\documentstyle[eqsecnum,psfig,aps,12pt]{revtex}
\def\beq{\begin{equation}}
\def\eeq{\end{equation}}

\def\bea{\arraycolsep .1em \begin{eqnarray}}
\def\eea{\end{eqnarray}}
\def\Tr{{\rm Tr}}
\def\step{\\[-1.5ex]}

\def\de{\delta}
\def\Ga{\Gamma}

\def\eq#1{(\ref{#1})}

\def\Eq#1{Eq.~(\ref{#1})}

\def\s0#1#2{\mbox{\small{$ \frac{#1}{#2} $}}}
\def\0#1#2{\frac{#1}{#2}}

\makeatletter

\renewenvironment{thebibliography}[1]
         {\section*{References}\frenchspacing\small
          \begin{list}{[\arabic{enumi}]}
         {\usecounter{enumi}\parsep=2pt\topsep 0pt
         \settowidth{\labelwidth}{[#1]}
         \leftmargin=\labelwidth\advance\leftmargin\labelsep
         \rightmargin=0pt\itemsep=0pt\sloppy}}{\end{list}}

\makeatother

\begin{document}
\begin{center}

\thispagestyle{empty}

{\normalsize\begin{flushright}CERN-TH-2002-023 \\[12ex] 
\end{flushright}}

\mbox{\large \bf Critical exponents from optimised 
renormalisation group flows}\\[6ex]

{Daniel F. Litim}
\\[2ex]
{\it Theory Division, CERN, CH -- 1211 Geneva 23.}
\\
{\footnotesize (Daniel.Litim@cern.ch)}
\\[10ex]
 
{\small \bf Abstract}\\[2ex]
\begin{minipage}{14cm}{\small 
    Within the exact renormalisation group, the scaling solutions for
    O($N$) symmetric scalar field theories are studied to leading
    order in the derivative expansion. The Gaussian fixed point is
    examined for $d>2$ dimensions and arbitrary infrared
    regularisation. The Wilson-Fisher fixed point in $d=3$ is studied
    using an optimised flow. We compute critical exponents and
    subleading corrections-to-scaling to high accuracy from the
    eigenvalues of the stability matrix at criticality for all $N$.
    We establish that the optimisation is responsible for the rapid
    convergence of the flow and polynomial truncations thereof. The
    scheme dependence of the leading critical exponent is analysed.
    For all $N\ge 0$, it is found that the leading exponent is
    bounded.  The upper boundary is achieved for a Callan-Symanzik
    flow and corresponds, for all $N$, to the large-$N$ limit. The
    lower boundary is achieved by the optimised flow and is closest to
    the physical value. We show the reliability of polynomial
    approximations, even to low orders, if they are accompanied by an
    appropriate choice for the regulator.  Possible applications to
    other theories are outlined.}
\end{minipage}
\end{center}

\newpage
\pagestyle{plain}
\setcounter{page}{1}
\noindent 
\section{Introduction}\label{Introduction}

Renormalisation group techniques are important tools to describe how
classical physics is modified by quantum fluctuations. Integrating-out
all quantum fluctuations provides the link between the classical
theory and the full quantum effective theory. Universality implies
that the details of the underlying classical theory -- other than the
global symmetries, long- or short-range interactions, and the
dimensionality -- are irrelevant for the characteristics of the quantum
effective theory.  For this reason, universal properties of phase
transitions in numerous physical systems (entangled polymers,
liquid-vapour transition, superfluid transition in ${}^4$He,
ferromagnetic transitions, QCD phase transition with two massless
quark flavours) can be addressed based on simple scalar field theories
\cite{Zinn-Justin:1989mi}.  \step
 
A useful method is given by the Exact Renormalisation Group (ERG)
\cite{Polchinski,continuum,Ellwanger:1994mw,Morris:1994qb,Bagnuls:2000ae},
which is based on the Wilsonian idea of integrating-out infinitesimal
momentum shells. The corresponding flow, which has a simple one-loop
structure, is very flexible concerning approximations, and its domain
of applicability is not tied to weak coupling.  Recently, it has been
shown that ERG flows can be optimised, thereby providing improved
results already to low orders within a given approximation
\cite{Litim:2000ci,Litim:2001up,Litim:2001fd,Litim:2001dt}. In the
present paper, we apply this idea to the universality class of $O(N)$
symmetric scalar theories in three dimensions and compute critical
exponents and subleading corrections to scaling.  We expect that
insights gained from this investigation will also prove useful for
applications to gauge theories \cite{Litim:1998nf} or gravity
\cite{QuantumGravity1}, which are more difficult to handle.\step

Universal critical exponents have been computed previously using
either polynomial truncations of exact renormalisation group flows, or
the derivative expansion to leading and subleading order
\cite{Litim:2001dt,Hasenfratz:1986dm,Margaritis:1988hv,Tetradis:1994ts,Morris:1994ie,Ball:1995ji,Litim:1995ex,Tetradis:1996br,Comellas:1997tf,Morris:1998xj,Liao:2000sh,VonGersdorff:2000kp,Litim:2001hk},
and in \cite{Litim:2001hk,Bohr:2001gp} based on the proper-time
renormalisation group \cite{Liao:1996fp}.  All results are affected by
the underlying approximations which induce a spurious dependence on
the regularisation
\cite{Litim:2001dt,Ball:1995ji,Liao:2000sh,Litim:2001ky,Litim:1997nw,Freire:2001sx}.
This is somewhat similar to the scheme dependence within perturbative
QCD, or within truncated solutions of Schwinger-Dyson equations. While
this scheme dependence should vanish at sufficiently high order in the
expansion, practical applications are always bound to a finite order,
and hence to a non-vanishing scheme dependence. In some cases, it has
even been observed that higher order results happen to be worse than
lower order ones \cite{Morris:1998xj}. In consequence, one should gain
some understanding of the spurious scheme dependence.  Without this,
it is difficult to decide which of the different scheme-dependent
results within a fixed truncation could be considered as trustworthy.
\step

A partial understanding of the interplay of approximations and scheme
dependence has been achieved previously. For scalar QED
\cite{ScalarQED}, the scheme dependence in the region of first order
phase transition has been studied in
\cite{Litim:1997nw,Freire:2001sx}. For $3d$ scalar theories, the
interplay between the smoothness of the regulator and the resulting
critical exponents has been addressed in \cite{Liao:2000sh} using a
minimum sensitivity condition. For Einstein quantum gravity, where a
new UV fixed point has been found recently to low orders in a
polynomial truncation, the corresponding analysis has been given in
\cite{QuantumGravity2}. The weak scheme dependence found in these
cases suggests that higher order corrections remain small, thereby
strengthening the results existing so far.  The evidence created this
way is partly circumstantial, because the range over which a physical
observable varies with the scheme depends on the class of regulators.
\step

Here, we address the problem from a different perspective. The main
ingredient in our analysis is the concept of optimisation
\cite{Litim:2000ci,Litim:2001up,Litim:2001fd,Litim:2001dt}.  For a
given physical problem, it should be possible to identify specific
regulators which lead to a better convergence behaviour of the flow.
This strategy is based only on the ERG flow itself and has lead to a
simple optimisation criterion for flows \cite{Litim:2000ci}. Optimised
flows have a number of interesting properties \cite{Litim:2001up}.
They lead to a fast decoupling of heavy modes, they disentangle
quantum and thermal fluctuations along the flow, and they lead to a
smooth approach towards a convex effective potential for theories in a
phase with spontaneous symmetry breaking.  The optimisation is closely
linked to a minimum sensitivity condition \cite{Litim:2001fd} in a
sense which will be made transparent below.  Furthermore, optimised
flows have been shown to improve the convergence of the derivative
expansion \cite{Litim:2001dt}. Thus, optimised flows are promising
candidates for high precision computations within this formalism.
Here, we do so within a local potential approximation.  \step

The second new ingredient of our analysis consists in a study of the
largest possible range of flows, and the corresponding critical
exponents. We find that the range is larger than
previously assumed.  Furthermore, the results from optimised flows are
located at the (lower) boundary and happen to be closest to the
physical values.  \step

For the numerical analysis, and apart from the local potential
approximation, we employ a polynomial approximation. This additional
approximation is reliable if it converges reasonably fast towards the
full solution.  However, it has been criticised previously in the
literature. For a sharp cut-off, it has lead to spurious solutions
\cite{Margaritis:1988hv}, and its convergence was found to be poor
\cite{Morris:1994ki}, which has lead to strong doubts concerning its
reliability (see also \cite{Aoki:1998um}).  In contrast to these
findings, we show that the poor convergence is an artifact of the
sharp cut-off regulator, rather than an artifact of the polynomial
approximation. Using an optimised flow, we find that the polynomial
approximation is stable and that it converges rapidly for all
technical purposes.  \step

The format of the paper is as follows. We review the basic ingredients
of the formalism and introduce the optimisation ideas
(Sect.~\ref{RGFlow}). Then, we introduce our numerical method and
study the non-trivial scaling solution in $3d$
(Sect.~\ref{FixedPoints}). We compute the eigenvalues at criticality
to high accuracy from an optimised flow. The convergence and stability
of the flow, and of the polynomial truncation, are established
(Sect.~\ref{CriticalExponents}). We study the scheme dependence of the
critical index $\nu$ (Sect.~\ref{Bounds}).  Finally, we discuss the
main results of the paper with particular emphasis on the predictive
power, on the convergence properties of flows, and on implications for
other theories (Sect.~\ref{Discussion}). Three appendices contain the
study of the Gaussian fixed point for arbitrary regularisation and
$d>2$ (App.~\ref{Gauss}), and technical details for specific classes
of regulators, including a computation of the corresponding flows and
critical exponents (Apps.~\ref{VariantsOpt} and~\ref{SlidingOpt}).

\section{RG flow for $O(N)$ symmetric scalar theories}\label{RGFlow}

Here, we briefly review some basic ingredients of the ERG formalism,
and its approximation to leading order in the derivative expansion. We
also discuss important aspects of the regularisation, and its
optimisation, which is employed in the following sections.

\subsection{Renormalisation group flows}

Exact renormalisation group equations are based on the Wilsonian idea
of integrating out momentum modes within a path integral
representation of quantum field theory \cite{Polchinski}.  In its
modern form, the ERG flow for an effective action $\Gamma_k$ for
bosonic fields $\phi$ is given by the simple one-loop expression
\cite{continuum,Ellwanger:1994mw,Morris:1994qb,Bagnuls:2000ae}
\beq \label{ERG}
\partial_t\Gamma_k[\phi] =
\frac{1}{2}\Tr\left( \frac{\de^2\Gamma_k}{\delta \phi\,\delta\phi}
                     + R_k\right)^{-1}    \partial_t  R_k
\eeq
Here, $t\equiv\ln k$ is the logarithmic scale parameter,
and $R_k(q^2)$ is an infrared (IR) regulator at the momentum scale
$k$.  From now on, we suppress the index $k$ on $R$. The flow
trajectory of \Eq{ERG} in the space of action functional depends on
the IR regulator function $R$. $R$ obeys a few restrictions, which
ensure that the flow equation is well-defined, thereby interpolating
between an initial action in the UV and the full quantum effective
action in the IR. We require that
\bea
\label{I}    \lim_{q^2/k^2\to 0}R(q^2)  >  0     \,,  \\
\label{II}   \lim_{k^2/q^2\to 0}R(q^2) \to 0     \,,  \\
\label{III}  \lim_{k\to\Lambda}R(q^2)  \to \infty\,.
\eea
Equation \eq{I} ensures that the effective propagator at vanishing
field remains finite in the infrared limit $q^2\to 0$, and no infrared
divergences are encountered in the presence of massless modes.
Equation \eq{II} guarantees that the regulator function is removed in
the physical limit, where $\Ga\equiv\lim_{k\to 0}\Ga_k$.  Equation
\eq{III} ensures that $\Ga_k$ approaches the microscopic action
$S=\lim_{k\to \Lambda}\Ga_k$ in the UV limit $k\to \Lambda$.  We put
$\Lambda=\infty$ in the sequel. For later use, we introduce a
dimensionless regulator function $r(y)$ as
\beq\label{r}   R(q^2)  =  q^2\, r(q^2/k^2)\,.  \eeq
Now we turn to the flow equation for an $O(N)$ symmetric scalar field
theory in $d$ dimensions to leading order in the derivative expansion,
the so-called local potential approximation \cite{Golner:1986}.  This
approximation amounts to the Ansatz
\beq\label{AnsatzGamma}
\Gamma_k=\int d^dx \left(U_k(\bar\rho) 
             + \012 \partial_\mu \phi^a\partial_\mu \phi_a
\right)\ .
\eeq
for the functional $\Gamma_k$. The Ansatz neglects higher order
corrections proportional to the anomalous dimension $\eta$ of the
fields. The latter are of the order of a few percent for the
physically interesting universality classes $N=0,\cdots, 4$. Hence, we
expect that a derivative expansion is sensible, and that the result of
a leading order computation is correct up to corrections of the order
of $\eta$. Inserting this Ansatz into \eq{ERG} and evaluating it for
constant fields leads to the flow for $U_k$. We rewrite this flow
equation in dimensionless variables $u(\rho)=U_k/k^d$ and $\rho=\s012
\phi^a\phi_a k^{2-d}$. In addition, the angular integration of the
momentum trace is performed to give \cite{Wetterich:1993be}
\beq\label{FlowPotential}
\partial_t u+du-(d-2) \rho u'
=2v_d(N-1)\ell(u')+2v_d \ell(2\rho u'')\,
\eeq
with $v^{-1}_d=2^{d+1}\pi^{d/2}\Gamma(\s0d2)$. The function
$\ell(\omega)$ are given by
\beq\label{Id}
\ell(\omega)
=\s012\int^\infty_0dy y^{d/2} \0{\partial_t r(y)}{y(1+r)+\omega}\,
\eeq
with $y\equiv q^2/k^2$ and $\partial_t r(y) = -2 y r'(y)$. The flow
\eq{FlowPotential} is a second order non-linear partial differential
equation. All non-trivial information regarding the renormalisation
flow and the regularisation scheme (RS) are encoded in the function
\eq{Id}. The momentum integration is peaked and regularised: for large
momenta due to the regulator term $\partial_t r(y)$, and for small
momenta due to $r(y)$ in the numerator. All terms on the left-hand
side of \Eq{FlowPotential} do not depend explicitly on the RS. They
simply display the intrinsic scaling of the variables which we have
chosen for our parametrisation of the flow.

\subsection{Optimisation}\label{Optimisation}

A good choice for the regulator is most important for a rapid
convergence and the stability of an approximated flow towards the
physical theory. Recently, is has been argued that such (optimised)
choices of the IR regularisation are indeed available
\cite{Litim:2000ci,Litim:2001up,Litim:2001fd}. The main observation is
that the flow trajectory of \eq{ERG} depends on the regularisation.
This observation is most important for approximated flows: typically,
their endpoint also depends spuriously on the regularisation. The
dependence is absent for the full integrated flow. Hence, in order to
provide reliable physical predictions, it is important to seek for
regularisations for which the main physical informations are already
contained within a few leading order terms of an approximation. This
issue is intimately linked to the stability of flows.  \step

The main ingredient in \eq{ERG} is the full inverse propagator.  Due
to the IR regularisation, the full inverse regularised propagator
displays a gap as a function of momenta
\cite{Litim:2001up,Litim:2001fd},
\beq\label{Effective1}
\min_{q^2\ge 0}
\left(
\left.
\0{\delta^{2}\Gamma_k[\phi]}{\delta \phi(q)\delta\phi(-q)}
\right|_{\phi=\phi_0}
+R_k(q^2)\right) = C\, k^2 >0\,.
\eeq 
The functional derivative is evaluated at a properly chosen expansion
point $\phi_0$. The existence of the gap $C>0$ implies an IR
regularisation, and is a prerequisite for the ERG formalism.
Elsewise, \eq{ERG} becomes singular at points where the full inverse
effective propagator develops zero modes.  It is expected that an
approximated ERG flow in the space of all action functionals is most
stable if the regularised full propagator is most regular along the
flow. This reasoning corresponds to maximising the gap. \step

To leading order in the derivative expansion, and dropping irrelevant
momentum-independent terms, the gap is given by
\beq\label{C}
C=\min_{y\ge 0} y(1+r(y))\,.
\eeq
Within the local potential approximation, the gap $C$ in \eq{C}
depends on the regularisation, but not on the specific theory
considered. A comparison of the gap of different regulators requires
an appropriate normalisation of $r$. In order to make the flow
\eq{ERG} more stable, we require that the gap, as a function of the
RS, is maximal in the momentum regime where the flow receives its main
contributions.  This is the optimisation criterion of
\cite{Litim:2000ci,Litim:2001up,Litim:2001fd}. It corresponds to an
optimisation of the radius of convergence of amplitude expansions and
the derivative expansion. It can also be shown that the optimisation
is closely linked to a minimum sensitivity condition.  Optimised
regulators are those for which the maximum in \eq{C} is attained. In
Ref.~\cite{Litim:2001up}, these considerations have lead to a very
specific solution to the optimisation condition, given by
\beq\label{Ropt} R_{\rm opt}(q^2) = (k^2-q^2)\,\theta (k^2-q^2)\,. \eeq
It has a number of interesting properties. For large momenta
$q^2>k^2$, the propagation of modes is fully suppressed since
$R\equiv 0$. For small momenta $q^2<k^2$, all modes propagate with
an effective mass term given by the IR scale, $q^2+R(q^2)=k^2$.
Based on the discussion in
Refs.~\cite{Litim:2000ci,Litim:2001up,Litim:2001fd}, we expect that
\Eq{Ropt} leads to an improved convergence and hence better physical
predictions already to leading order in the derivative expansion.
Rewriting \eq{Ropt} in the form of \eq{r} leads to
\beq\label{ropt} r_{\rm opt}(y) = (\s01y-1)\,\theta (1-y) \eeq 
Below, we mainly employ this regulator, and variants thereof (cf.
Appendices~\ref{VariantsOpt} and \ref{SlidingOpt}). When expressed in
terms of the optimised regulator \eq{Ropt}, the flow equation
\eq{FlowPotential} becomes
\beq\label{FlowPotentialOpt}
\partial_t u=   -d u 
                +(d-2) \rho u'
                +\04d v_d \0{N-1}{1+u'} 
                +\04d v_d \0{1}{1+u'+2\rho u''}\,.
\eeq
Notice that the numerical factors $(4v_d)/d$ can be absorbed into the
potential and the fields by an appropriate rescaling.

\section{Fixed points}\label{FixedPoints}

The flow equation \eq{FlowPotentialOpt} is known to exhibit two
scaling solutions in $2<d<4$, which correspond to the Gaussian and the
Wilson-Fisher fixed point, respectively. The Gaussian fixed point is
given by the trivial solution $u_\star=$ const, and is discussed in
the Appendix~\ref{Gauss} for arbitrary regularisation and $2<d<4$.
In this section, we study the non-trivial fixed point $u_\star\neq$
const.~in $3d$ based on an optimised flow. We introduce the numerical
method, and discuss the scaling form of the fixed point solution.
Universal critical exponents are computed in the next section.
 
\subsection{Numerics}

Numerical methods for solving partial differential equations are
well-known. Here, we employ a polynomial truncation of the scaling
potential, retaining vertex functions $\phi^{2 n}$ up to a maximum
number $n_{\rm trunc}$. Polynomial approximations are reliable if they
depend only very weakly on higher order operators beyond some finite
order of the truncation.  (In the following section, we explicitly
confirm that this is indeed the case.)  Two different polynomial
expansions of the potential are used: expansion I corresponds to
\beq\label{PolyAnsatz0}
u(\rho)=\sum^{n_{\rm trunc}}_{n=1} \01{n!} \lambda_n \rho^n\,.
\eeq
In \Eq{PolyAnsatz0}, $\lambda_1$ denotes the (dimensionless) mass term
at the origin. We have normalised the potential as $u(\rho=0)=0$.  All
higher order coefficients $\lambda_n$ denote the $n$-th order coupling
at vanishing field. Expansion II, alternatively, approximates the
potential about the potential minimum $\rho=\rho_0$ as
\beq\label{PolyAnsatz}
u(\rho)=\sum^{n_{\rm trunc}}_{n=2} \01{n!} \lambda_n (\rho-\lambda_1)^n\,.
\eeq
In \Eq{PolyAnsatz}, $\lambda_1$ denotes the location of the potential
minimum defined by $u'(\rho=\lambda_1)=0$. We have normalised the
potential as $u(\lambda_1)=0$. All higher order coefficients
$\lambda_n$ denote the $n$-th order coupling at the potential minimum.
Both expansions \eq{PolyAnsatz0} and \eq{PolyAnsatz} are symmetric
under $\phi\to -\phi$, and approximate the potential as an even
polynomial in $\phi$. The number of independent operators contained in
\Eq{PolyAnsatz0} or \Eq{PolyAnsatz} is $n_{\rm trunc}$.  The
expansions transform the partial differential equation
\eq{FlowPotential} into $n_{\rm trunc}$ coupled ordinary differential
equations $\partial_t \lambda_i \equiv \beta_i(\{\lambda_n\})$ for the
set of couplings
\beq
\{\lambda_n,1\le n\le n_{\rm trunc}\}\,.
\eeq
For the numerical study, we use mostly expansion II, which is known to
have superior convergence properties in comparison to expansion I
\cite{Aoki:1998um,Litim:2000ci}.

\subsection{Wilson-Fisher fixed point}

The Wilson-Fisher fixed point corresponds to the non-trivial scaling
solution of \eq{FlowPotentialOpt}. Here, we restrict ourselves to the
case $d=3$, and to the optimised regulator as introduced above. The
scaling potential obeys the differential equation
\beq\label{FlowPotential3}
         0  =   -3 u _\star
                + \rho u'_\star
                +\01{6\pi^2} \0{N-1}{1+u'_\star} 
                +\01{6\pi^2} \0{1}{1+u'_\star+2\rho u''_\star}+{\rm const.}\,
\eeq
and $u _\star(\rho)\neq{\rm const}$. The constant in
\Eq{FlowPotential3} can be fixed freely, and we chose it such that the
scaling potential vanishes at its minimum, $u_\star=0$ at the point
$\rho_0$ with $u'_\star=0$. An analytical solution of
\Eq{FlowPotential3} has been given in Ref.~\cite{Litim:1995ex} for the
limit $N=\infty$. For $N\neq\infty$, and in the vicinity of $\rho=0$,
the scaling solution can be obtained analytically as a Taylor
expansion in the field. For $d=3$ and $N=1$, and inserting the
expansion \eq{PolyAnsatz0} into \eq{FlowPotential3} with const.~$=0$,
we find
\begin{mathletters}\label{solutionExpansion}
  \bea
\lambda_0&=&\ \ \s01{18}\pi^{-2}(1 + \lambda_1)^{-1}\\
\lambda_2&=&-4\pi^2\,\lambda_1\,{\left( 1 + \lambda_1 \right) }^2 \\
\lambda_3&=&\ \
\s0{72}{15}\pi^4\,\lambda_1\, {\left( 1 + \lambda_1 \right) }^3\,
                               \left( 1 + 13\,\lambda_1 \right) \\
\lambda_4&=& 
-\s0{1728}{7}\pi^6\,\lambda_1^2\,{\left( 1 + \lambda_1\right) }^4\,
  \left( 1 + 7\,\lambda_1 \right) \\
\lambda_5&=&\ \
\s0{768}{7}\pi^8\,\lambda_1^2\,{\left( 1 + \lambda_1 \right) }^5\,
                          ( 2 + 121\,\lambda_1 + 623\,\lambda_1^2)\\
  &\vdots&\nonumber \eea
\end{mathletters}%
for small fields. Similar explicit solutions are found for $d\neq 3$
or $N\neq 1$. Notice that all couplings are expressed as functions of
the dimensionless mass term at vanishing field, $\lambda_1$. The
domain of validity of this expansion is restricted to $0\le \rho\le
\rho_c<\infty$. From the explicit solution for large fields,
\eq{solutionExpansionLarge} below, it is clear that the polynomial
approximation cannot be extended to arbitrary large fields.  The value
$\rho_c$ defines the radius of convergence for the polynomial
approximation. It is linked to the gap parameter $C$ introduced
earlier \cite{Litim:2000ci}. Not all values for $\lambda_1$ lead to a
scaling solution which remains finite and analytical for all
$\rho<\rho_c$. It is at this point where the quantisation of
$\lambda_1$ becomes manifest: only two values for $\lambda_1$
correspond to well-defined solutions of the fixed point equation,
given by the Gaussian fixed point $\lambda_1=0$ and the Wilson-Fisher
fixed point $\lambda_1=\lambda_{1\star}< 0$.  The value for
$\lambda_{1\star}$ is determined by fine tuning $\lambda_1$ such that
the solution \eq{solutionExpansion} extends to $\rho=\rho_c$. The
result is $\lambda_{1\star} = -0.1860642\cdots$ for $N=1$.  In the
other limit, for large fields $\rho\gg 1$, we find
\beq\label{solutionExpansionLarge}
u(\rho)=A\,\rho^3+\s0{1}{450\pi^2}\,A^{-1}\,\rho^{-2} +\ldots\,,
\eeq
(and similarly for $N\neq 1)$, where the dots denote subleading terms
in $\rho$ and the constant $A>0$ has to be fixed appropriately, in a
way similar to $\lambda_1$ in \eq{solutionExpansion}. It would be
interesting to study the analyticity properties of the fixed point
solutions more deeply, and to contrast it with the analysis of
\cite{Morris:1994ki} for the sharp cut-off flow.  Despite the simple
explicit form of the flow, and its explicit solution for large and
small fields, it is more efficient to identify the scaling solution
and the related critical exponents using numerical methods.

\subsection{Scaling potential}

We have solved the resulting coupled set of differential equations in
$d=3$ dimensions, for $N=0,1,\cdots, 10$ and for truncations up to
$n_{\rm trunc}=20$. We have retained only the solution which has one
unstable eigendirection, corresponding to the Wilson-Fisher fixed
point. Fig.~1 shows the corresponding scaling potential $u_\star$ in
the vicinity of the potential minimum. In Fig.~1, the unusual
normalisation of the potential has been chosen only for display
purposes. The scaling potential has, for all $N$, a minimum at
non-vanishing field. For this reason, the expansion II has better
convergence properties than expansion I.

\begin{figure}
\begin{center}
\vskip-1cm
\unitlength0.001\hsize
\begin{picture}(1000,550)
\put(290,155){\framebox{\large $ c_N\, u_\star(\rho)$}}
\put(220,80){\large $\rho$}
\put(740,85){\large $\rho$}
\put(840,160){\framebox{\large $u'_\star(\rho)$}}
\psfig{file=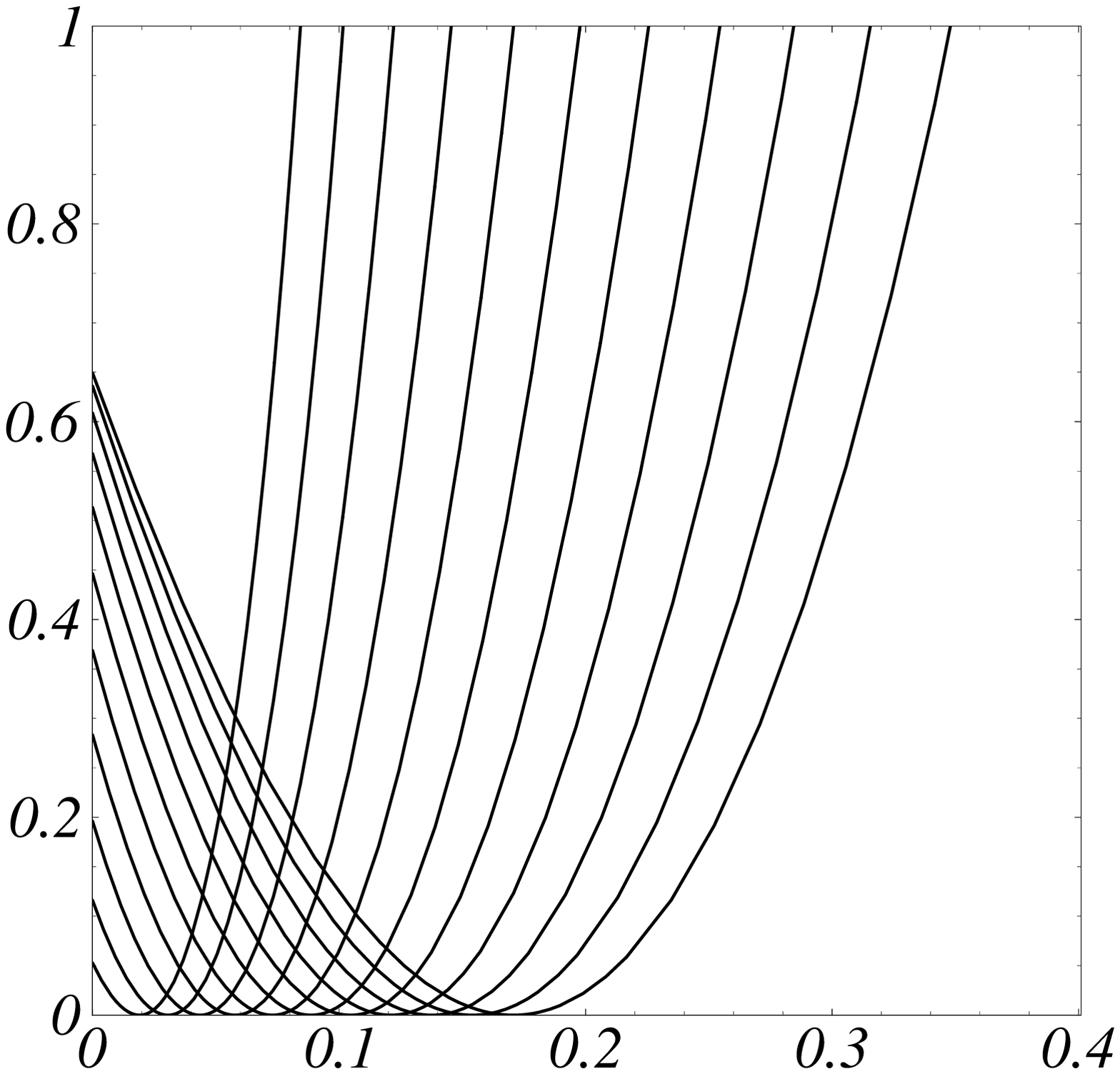,width=.45\hsize}
\hskip.05\hsize
\psfig{file=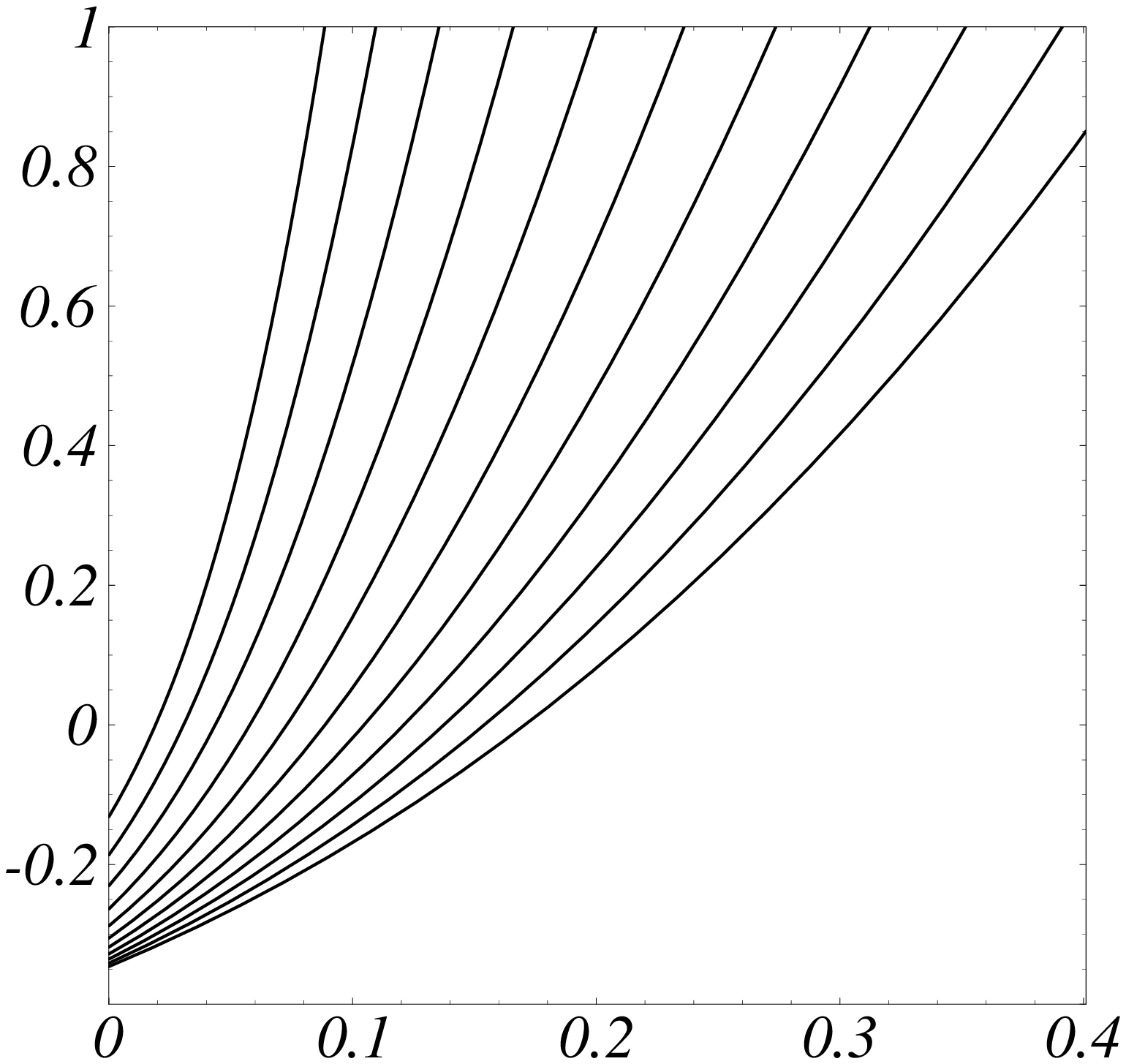,width=.45\hsize}
\end{picture}
\vskip-.75cm
\begin{minipage}{.47\hsize}
  { \small {\bf Figure 1:} The scaling potential, with $c_N=40-2N$.
    From left to right: $N=0,1,\cdots,10$.}
\end{minipage} 
\hskip.05\hsize
\begin{minipage}{.47\hsize}
  { \small {\bf Figure 2:} The amplitude $u'_\star(\rho)$ of the scaling
  solution. From left to right: $N=0,1,\cdots,10$.}
\end{minipage} 
\end{center}
\end{figure}

\begin{figure}
\begin{center}
\vskip-1cm
\unitlength0.001\hsize
\begin{picture}(1000,550)
\put(190,150){\framebox{\large $u'_\star(\rho)+2\rho u''_\star(\rho)$}}
\put(220,70){\large $\rho$}
\hskip.05\hsize
\put(520,150){
\begin{minipage}{.47\hsize}
  {\small {\bf Figure 3:} The amplitude $u'_\star+2\rho u''_\star$ of
    the scaling solution. The dots indicate the points where
    $u'_\star=0$. From left to right: $N=0,1,\cdots,10$.}
\end{minipage}} 
\psfig{file=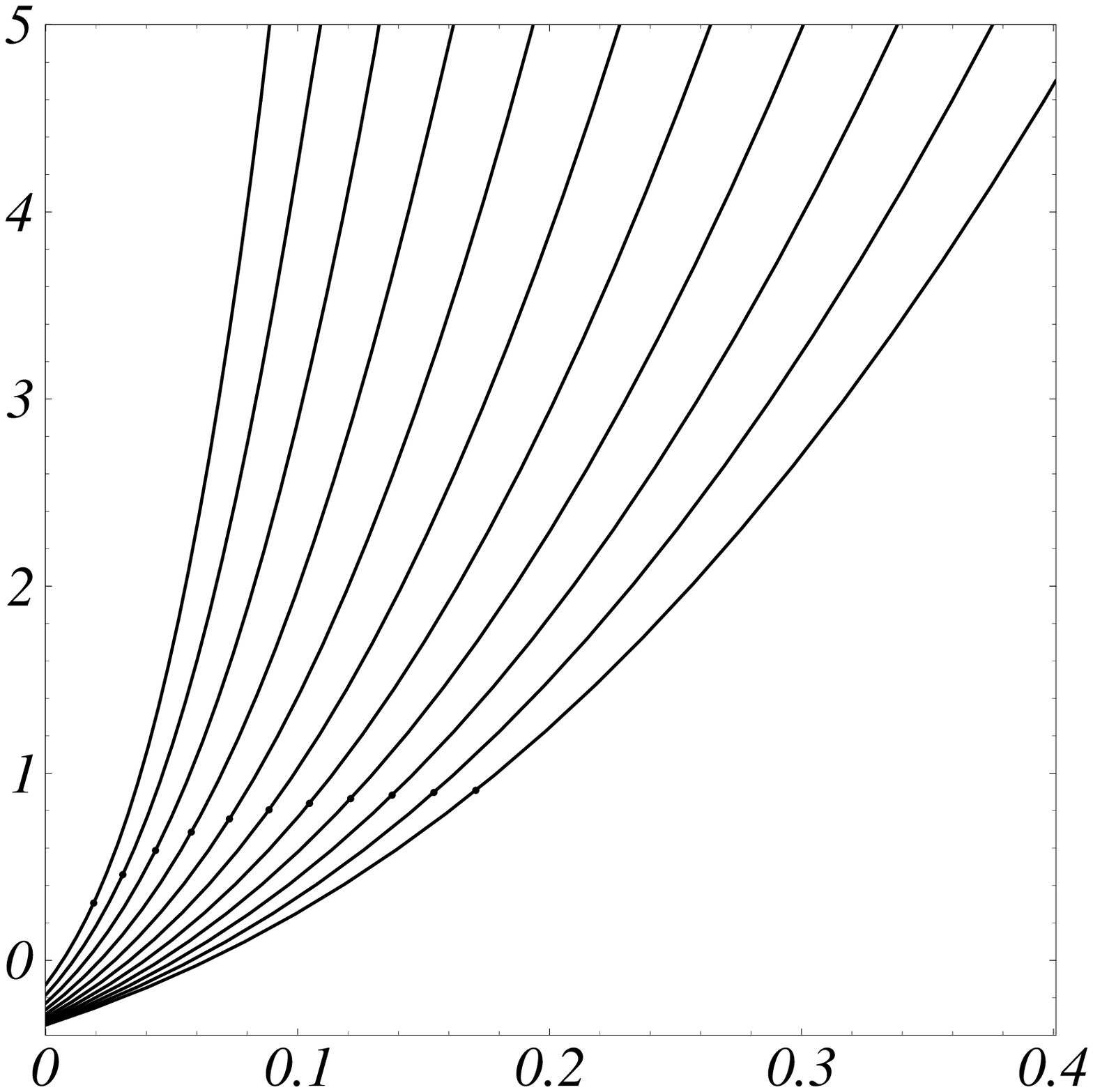,width=.45\hsize}
\end{picture}
\end{center}
\vskip-1.cm
\end{figure}

Fig.~2 shows the amplitude $u'$ at the fixed point. It is a monotonic
function of the field. It approaches its most negative values at
vanishing field. The flow equation would have a pole at points where
$1+u'$ vanishes. This is, however, never the case, because $1+u'$
stays always positive. Fig.~3 shows the radial mode of the scaling
potential $u'_\star+ 2\rho u''_\star$ in the vicinity of the potential
minimum. It is a monotonic function of the fields. The dots in Fig.~3
indicate where the derivative $u'_\star$ changes sign. Again, the most
negative value is attained at vanishing field and decreases with
increasing $N$, but it stays always above the pole of the flow
equation, $1+u'_\star+ 2\rho u''_\star>0$.\step

The scaling solution is non-universal.  However, critical exponents or
the eigenvalues of small perturbations about the scaling solutions
{\it are} universal. For their determination, it is sufficient to
study the flow of small perturbations in the vicinity of the scaling
potential, which is done next.

\section{Critical exponents}\label{CriticalExponents}

In this section, we compute the critical exponents to leading order in
the derivative expansion. Numerical results are given up to six
significant figures (Tab.~1). A higher precision can be achieved, but
is not required at the present state. We find a rapid convergence of
the polynomial approximation, for all observables considered. Our
results are compared with those from other regulators.

\subsection{Eigenvalues}

Given the Wilson-Fisher fixed point solution, we can seek for
universal critical exponents. In the vicinity of the non-trivial fixed
point, we have to solve the eigenvalue equation
\beq\label{WFFP:1}
\partial_t\ \delta u^{(m)}=\omega\,\delta u^{(m)}
\eeq
in order to determine the various universal eigenvalues $\omega$.
Using the flow equation, setting $d=3$ and choosing $m=1$, and
expanding the flow to leading order about the Wilson-Fisher fixed
point, we find
\bea
0&=&
\left[\omega+2-\0{N}{6\pi^2}\0{u''_\star}{(1+u'_\star)^2}\right]
\,\delta u' 
+\,\0{1}{3\pi^2}
\left[ \0{2N}{1+u'_\star}
-\0{1+u'_\star-\rho u''_\star-2\rho^2 u'''_\star}
   {(1+u'_\star+2\rho u''_\star)^2}\right] 
\,\delta u''
\nonumber \\ &&
+\,\0{1}{3\pi^2}\,\0{\rho}{1+u'_\star+2\rho u''_\star}\ \delta u'''\,.
\label{WFFP:2}
\eea
Instead of solving \Eq{WFFP:2} directly for an eigenperturbation
$\delta u'$ with eigenvalue $\omega$, we follow a slightly different
line which is numerically less demanding. Based on the
polynomial expansion used in the previous section, the fixed point
potential is given by the set of couplings $\{\lambda_{n,\star}\}$. At
the fixed point, the flow of the couplings $\partial_t\lambda_n\equiv
\beta_n$ vanishes, $\beta_n(\lambda_{i,\star})=0$. The eigenvalues
$\omega$ of the stability matrix at criticality
\beq\label{M}
M_{ij}=
\left.\partial\beta_i/\partial\lambda_j\right|_{\lambda=\lambda_{\star}}
\eeq
correspond to the eigenvalues of \Eq{WFFP:1}. Hence, the problem of
finding the eigenvalues of \Eq{WFFP:1} reduces to the problem of
finding the eigenvalues of the stability matrix $M$.\step

\begin{center}
\begin{tabular}{c|c|c|c|c|c}
$\quad{}     N   \quad{}    $&
$\quad\quad\quad{}\nu     \quad\quad\quad{}$&
$\quad\quad\quad{}\omega\quad\quad\quad{}$&
$\quad\quad\quad{}\omega_2\quad\quad\quad{}$&
$\quad\quad\quad{}\omega_3\quad\quad\quad{}$&
$\quad\quad\quad{}\omega_4\quad\quad\quad{}$
\\[.5ex] \hline
0
&0.592083
&0.65788
&3.308
&6.16
&9.2
\\
1
&0.649562
&0.655746
&3.180
&5.912
&8.80
\\
2
&0.708211
&0.671221
&3.0714
&5.679
&8.440
\\
3
&0.761123
&0.699837
&2.9914
&5.482
&8.125
\\
4
&0.804348
&0.733753
&2.9399
&5.330
&7.867
\\
5
&0.837741
&0.766735
&2.9108
&5.2195
&7.665
\\
6
&0.863076
&0.795815
&2.8967
&5.1409
&7.512
\\
7
&0.882389
&0.820316
&2.8916
&5.0863
&7.396
\\
8
&0.897338
&0.840612
&2.89163
&5.04848
&7.3086
\\
9
&0.909128
&0.857384
&2.89438
&5.02232
&7.2425
\\
10
&0.918605
&0.871311
&2.89846
&5.00420
&7.1921
\\
$\infty$
&1
&1
&3
&5
&7
\end{tabular}
\end{center}
\begin{center}
\begin{minipage}{\hsize}
  \vskip.3cm {\small {\bf Table 1:} The first five eigenvalues
    $\omega_0<0<\omega_1<\omega_2<\omega_3<\omega_4$, with
    $\nu=-1/\omega_0>0$ for various $N$ and $d=3$ dimensions.}
\end{minipage}
\end{center}

\subsection{Results}

We have computed the eigenvalues of \eq{M} as functions of the order
of the polynomial truncation. The numerical results for the first five
eigenvalues for various $N$ and $d=3$ dimensions are given in Tab.~1
and in Figs.~4 -- 8. Results for other regulators are compared in
\cite{Litim:2001dt} (see also Sects.~\ref{SectionScheme}
and~\ref{Bounds}), and results for the asymmetric corrections to
scaling are given elsewhere. Exact results, independent on the
regularisation, are known for $N=\infty$. The exact eigenvalues are
given by $\omega_n=2n-1 +{\cal O}(1/N)$, and only the subleading
corrections ${\cal O}(1/N)$ depend on the regulator. For $N=-2$, it is
known that $\nu=\s012$. Tab.~1 shows the first five eigenvalues,
ordered as $\omega_0<0<\omega<\omega_2<\omega_3<\omega_4$. The
critical exponent $\nu$ is given by $\nu=-1/\omega_0>0$. As a function
of $N$, the eigenvalue $\nu$ interpolates monotonically between the
exact values $\nu=\s012$ for $N=-2$ and $\nu=1$ for $N=\infty$. The
eigenvalues $\omega$ and $\omega_2$ are non-monotonic functions of
$N$. For small $N$, $\omega$ decreases until $N\approx 1$, and
increases towards the exact asymptotic value $\omega=1$ for
$N\to\infty$.  The eigenvalue $\omega_2$ decreases until $N\approx
7-8$ before settling to the asymptotic value $\omega_2=3$ at
$N=\infty$. The eigenvalues $\omega_3$ and $\omega_4$ are
monotonically decreasing functions of $N$.

\begin{figure}[t]
\begin{center}
\vskip-.8cm
\unitlength0.001\hsize
\begin{picture}(1000,550)
\put(330,400){\framebox{\large $\nu(N)$}}
\put(210,70){\large $n_{\rm trunc}$}
\put(830,460){\framebox{\large $\omega(N)$}}
\put(710,70){\large $n_{\rm trunc}$}
\psfig{file=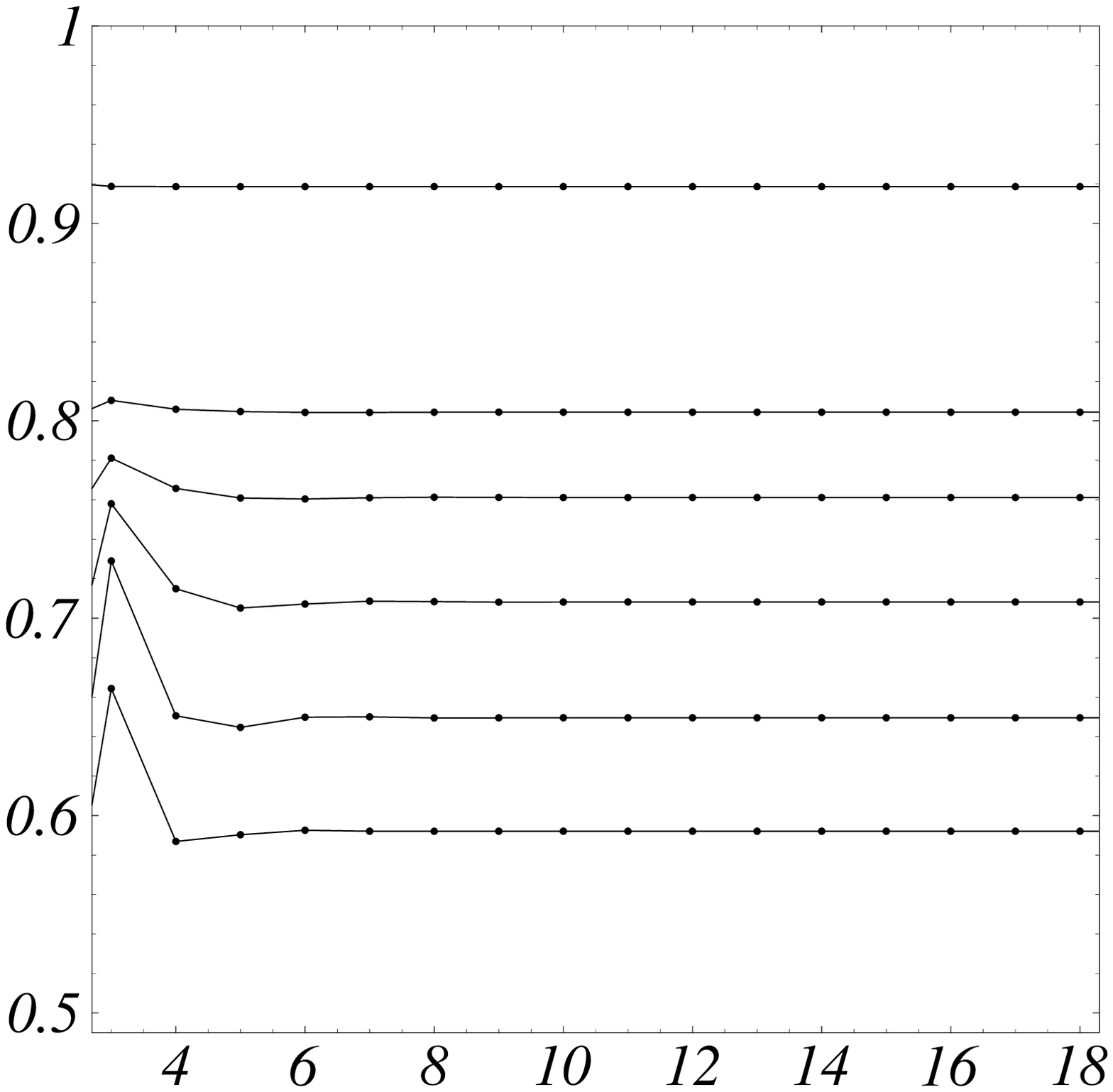,width=.45\hsize}
\hskip.05\hsize
\psfig{file=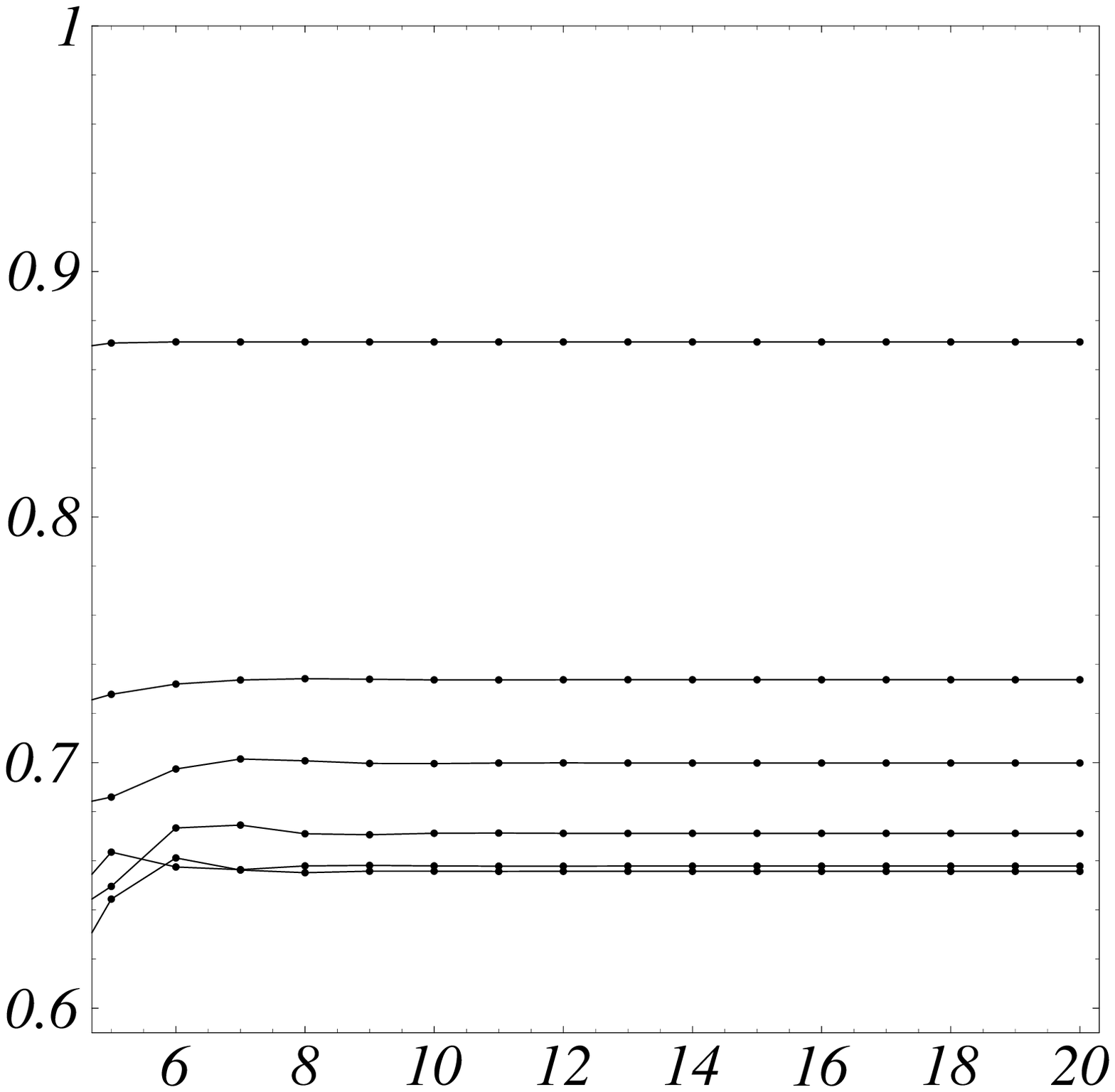,width=.45\hsize}
\end{picture}
\vskip-.75cm
\begin{minipage}{.45\hsize}
{\small {\bf Figure 4:} The exponent $\nu(N)$ as a function
  of $N$ and of the order of the truncation. From top to bottom:
  $N=10,4,3,2,1,0$.}
\end{minipage} 
\hskip.05\hsize
\begin{minipage}{.45\hsize}
{\small {\bf Figure 5:} The eigenvalue $\omega(N)$ as a function of
  $N$ and of the order of the truncation. From top to bottom:
  $N=10,4,3,2,0,1$.}
\end{minipage} 
\end{center}
\vskip-.1cm
\end{figure}

Fig.~4 shows the results for $\nu$ as a function of $n_{\rm trunc}$
and $N$. It is seen that the expansion is very stable. It convergences
already within low order of the truncation towards the asymptotic
value. Typically, $n_{\rm trunc}\approx 6$ gives the correct result
below the percent level. Furthermore, the convergence is better for
larger values of $N$. This behaviour is observed for all eigenvalues
(e.g.~Figs.~4 -- 8). Fig.~5 shows the same behaviour for the smallest
subleading eigenvalue $\omega$. For $n_{\rm trunc}\approx 8$, it has
settled below the percent level of the correct result.
\step

\begin{figure}[t]
\begin{center}
{}\vskip-2cm
\unitlength0.001\hsize
\begin{picture}(1000,550)
\put(330,470){\framebox{\large $\omega_2(N)$}}
\put(210,70){\large $n_{\rm trunc}$}
\put(830,470){\framebox{\large $\omega_3(N)$}}
\put(710,70){\large $n_{\rm trunc}$}
\psfig{file=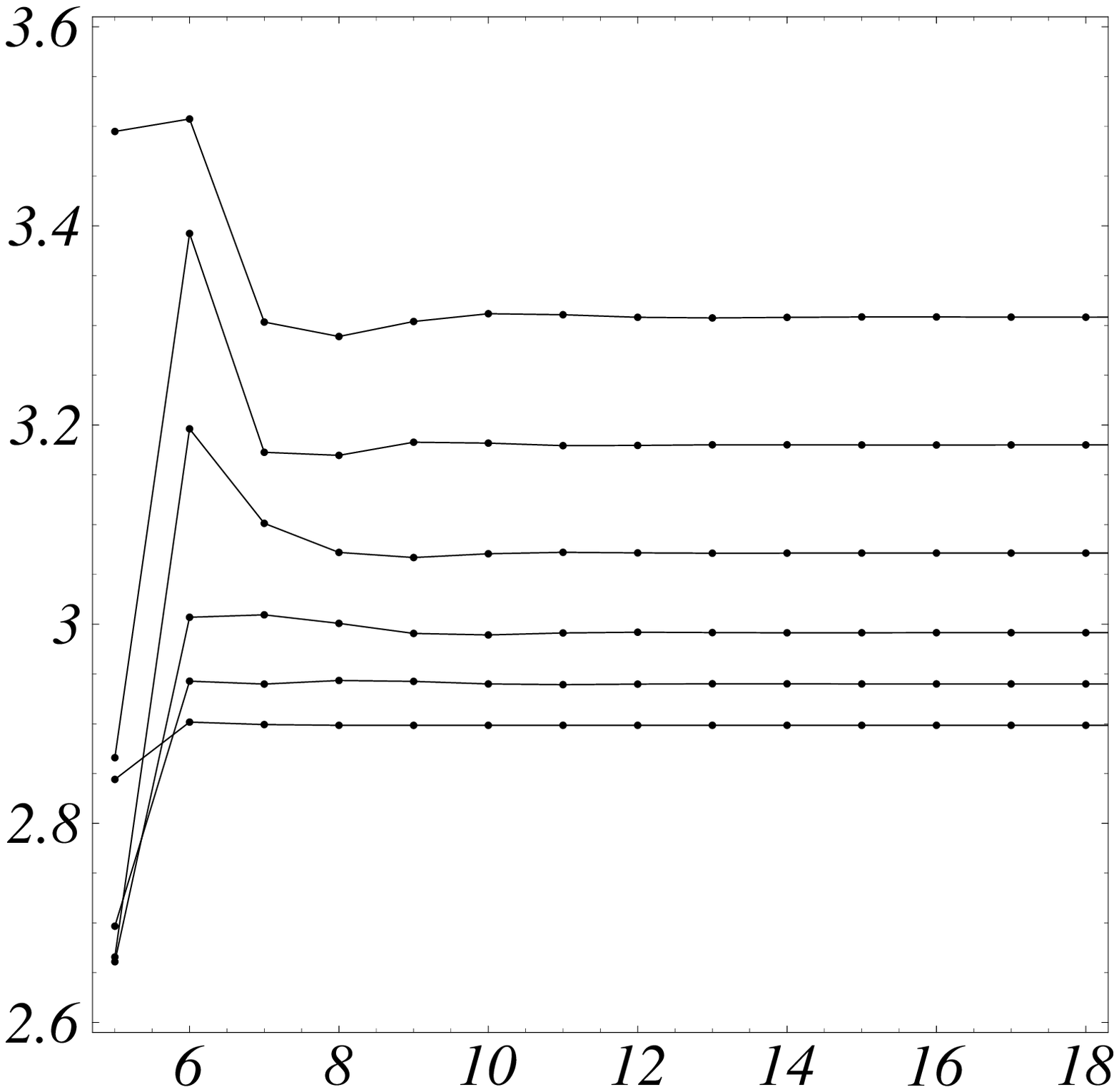,width=.45\hsize}
\hskip.05\hsize
\psfig{file=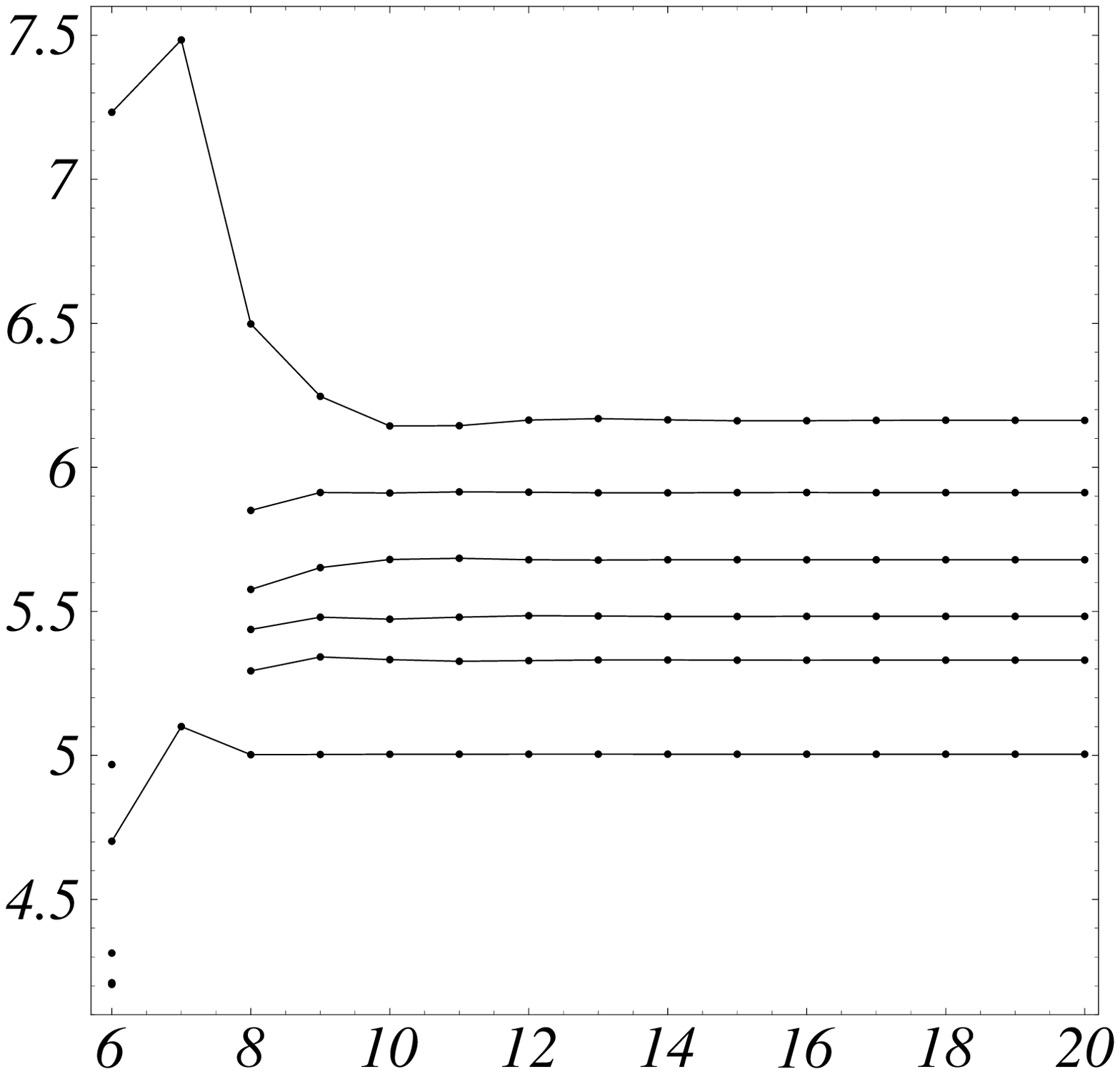,width=.45\hsize}
\end{picture}
\vskip-.75cm
\begin{minipage}{.47\hsize}
{\small {\bf Figure 6:} The critical index $\omega_2(N)$ as a
  function of $N$ and of the order of the truncation. From top to
  bottom: $N=0,1,2,3,4,10$.}
\end{minipage} 
\hskip.05\hsize
\begin{minipage}{.47\hsize}
{\small {\bf Figure 7:} The critical index $\omega_3(N)$ as a
  function of $N$ and of the order of the truncation. From top to
  bottom: $N=0,1,2,3,4,10$.}
\end{minipage} 
\end{center}
\vskip-1.cm
\end{figure}

\begin{figure}
\begin{center}
\unitlength0.001\hsize
\begin{picture}(1000,550)
\put(340,470){\framebox{\large $\omega_4(N)$}}
\put(240,70){\large $n_{\rm trunc}$}
\put(520,220){
\begin{minipage}{.47\hsize}
{\small {\bf Figure 8:} The critical index $\omega_4(N)$ as a
  function of $N$ and of the order of the truncation. From top to
  bottom: $N=0,1,2,3,4,10$. The isolated point at $n=7$ corresponds to
  $N=10$, and the two at $n=9$ to $N=2$ (upper) and $N=3$ (lower). The
  intermediate points (at $n=8$ and $n=10$, resp.) are missing because
  they have a small imaginary part.}
\end{minipage}}
\hskip.02\hsize
\psfig{file=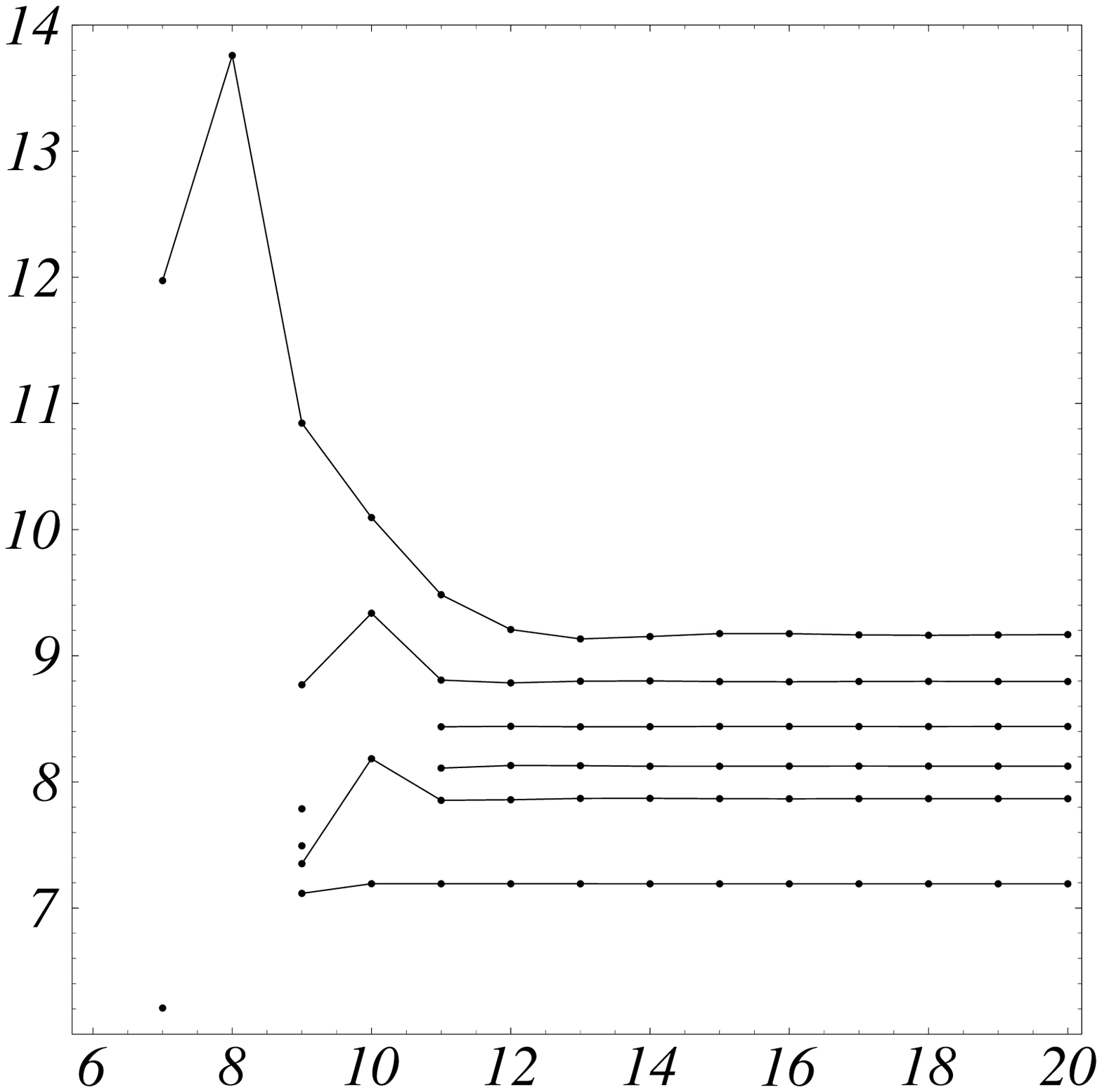,width=.45\hsize}
\end{picture}
\end{center}
\vskip-1.5cm
\end{figure}

For low orders of the truncation, the eigenvalues $\omega_2$,
$\omega_3$ and $\omega_4$ (Figs.~6, 7 and 8, respectively) depend more
strongly on $n_{\rm trunc}$, and do not even lead to real eigenvalues
in some cases (c.f.~Figs.~7 and 8). Again, the dependence is even
stronger for smaller $N$.  For sufficiently high order in the
truncation, however, all eigenvalues are real, and the convergence
towards the asymptotic value is fast. Typically, $\omega_2$,
$\omega_3$ and $\omega_4$ reach their asymptotic values below the
percent level for $n_{\rm trunc}\approx 10, 12$ and $14$. \step

For all eigenvalues, the basic picture is the same: for small $n_{\rm
  trunc}$, the truncation tends to overshoot the asymptotic value, but
with increasing $n_{\rm trunc}$ it relaxes towards it with a remaining
oscillation and decreasing amplitude.  From a specific order onwards,
the truncation sits -- for all technical purposes -- on top of the
asymptotic value. For a fixed level of accuracy, a lower order in the
truncation is required for the dominant observables like $\nu$ or
$\omega$, while a higher order is required for the subleading
eigenvalues.  Roughly speaking, to obtain the eigenvalue $\omega_n,
n=0,\cdots$ accurate below the percent level, a truncation with
$n_{\rm trunc}\approx 2n+6$ independent couplings is required.  

\subsection{Convergence and stability}

Next, we discuss the convergence and stability of the polynomial
approximation for an optimised flow.  From the results presented so
far (cf.~Figs.~4 - 8), we conclude that the optimised flow
\eq{FlowPotentialOpt} leads to a fast convergence of the polynomial
approximation for the scaling potential. More importantly, we have
seen that the inclusion of further vertex functions --- increasing
$n_{\rm trunc}\to n_{\rm trunc}+1$ --- does not alter the fixed point
structure. Rather, it leads to a small modification of the actual
fixed point solution and to minor corrections for the critical
exponents. This implies that the flow is very stable, and that most of
the physical information is already contained in a few leading order
terms of the truncation. This result is by no means trivial. A counter
example is furnished by the sharp cutoff (see also the following
section), where the convergence of the polynomial approximation is
poor. Here, the good convergence hinges on the use of an appropriately
optimised regulator
\cite{Litim:2000ci,Litim:2001up,Litim:2001fd}.

\begin{figure}
\begin{center}
{}\vskip-2.cm
\unitlength0.001\hsize
\begin{picture}(1000,550)
\put(200,360){ \fbox{{ $\displaystyle 
\log_{10}\left|\0{\nu_{\rm opt}-\nu_{\rm trunc}}{\nu_{\rm opt}}\right|$}}}
\put(230,-10){\large $n_{\rm trunc}$}
\put(520,140){
\begin{minipage}{.47\hsize}
  { \small {\bf Figure 9:} Ising universality class.  Convergence of
    $\nu_{\rm trunc}$ (expansion II) towards $\nu_{\rm opt}$ with
    increasing truncation. Points where $\nu_{\rm trunc}$ is larger
    (smaller) than $\nu_{\rm opt}$ are denoted by o ($\bullet$).
    Roughly speaking, for $2<n_{\rm trunc}<20$, the accuracy of the
    critical exponents improves by one decimal point every $\Delta
    n\approx 2-2.5$.  }
\end{minipage}} 
\hskip.02\hsize
\psfig{file=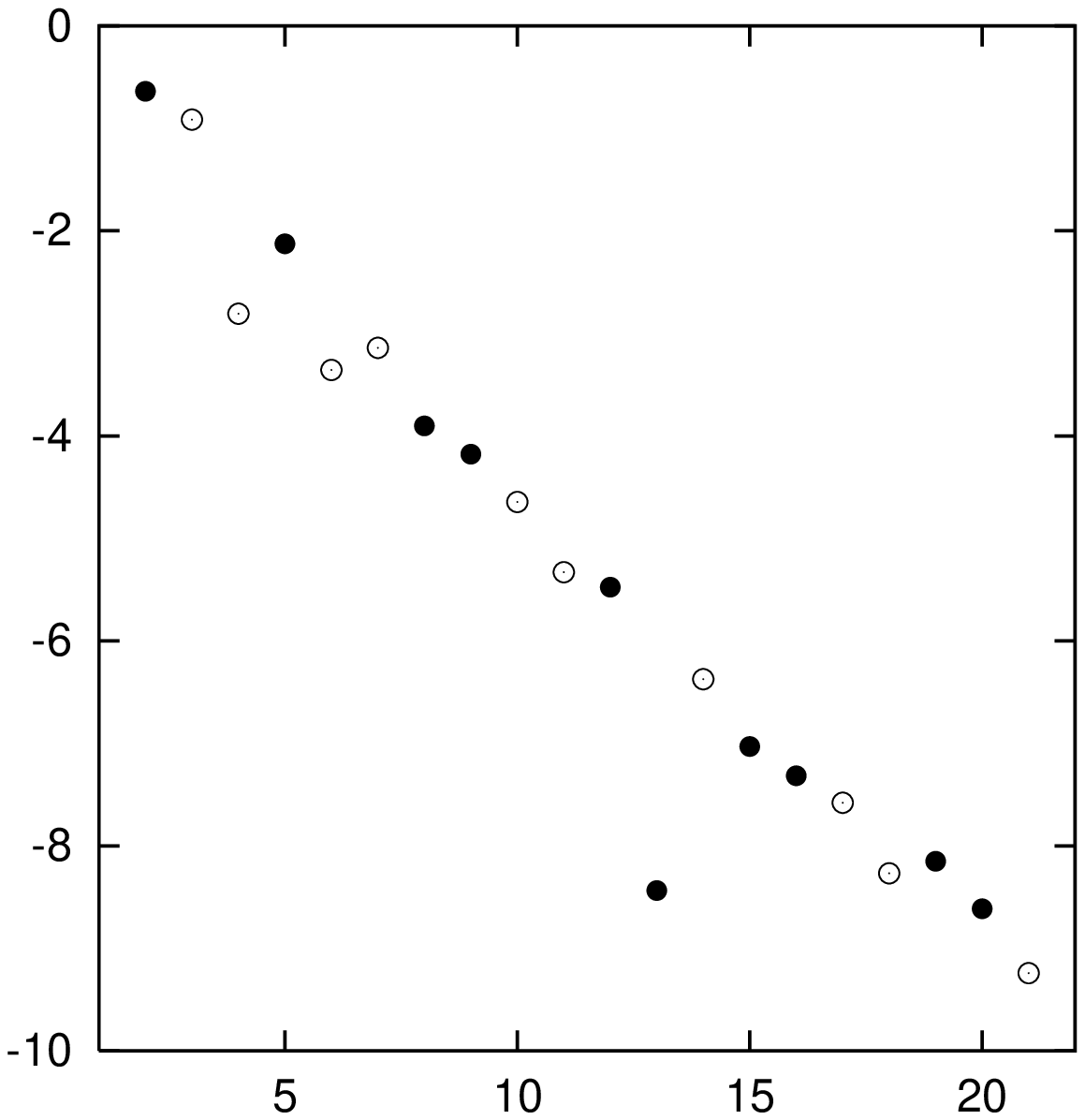,width=.45\hsize}
\end{picture}
\end{center}
\end{figure}

Let us have a closer look at the rate of convergence towards the
asymptotic values of expansion II (cf.~Fig.~9).  We denote with
$\nu_{\rm trunc}$ the approximate critical exponent which retains
$n=n_{\rm trunc}$ independent parameters in the effective action. The
semi-logarithmic plot in Fig.~9 then shows the rate with which
successive approximations converge towards the asymptotic value
$\nu_{\rm opt}$.  The series $\nu_{\rm trunc}$ oscillates about the
asymptotical values with a decreasing amplitude and, roughly, a
four-fold periodicity in the pattern $++--$.  The curve in Fig.~9 can
be approximated by a straight line with a slope $\approx -.4$ to
$-.5$. Hence, for every $\Delta n\approx 2 - 2.5$, the accuracy of
$\nu_{\rm trunc}$ increases by one decimal place.  Here, we have
analysed the convergence for $N=1$. For larger $N$, the convergence is
typically faster than for $N=1$, while for $N=0$, it is about the
same. Hence, the present considerations are qualitatively the same for
all $N$.

\begin{figure}
\begin{center}
\vskip-.5cm
\unitlength0.001\hsize
\begin{picture}(850,750)
\put(200,350){\large $n_{\rm trunc}$}
\put(610,350){\large $n_{\rm trunc}$}
\put(280,450){\fbox{\large $\nu_{\rm Ising}$}}
\put(110,510){I}
\put(110,650){II}
\put(600,650){I}
\put(520,550){II}
\psfig{file=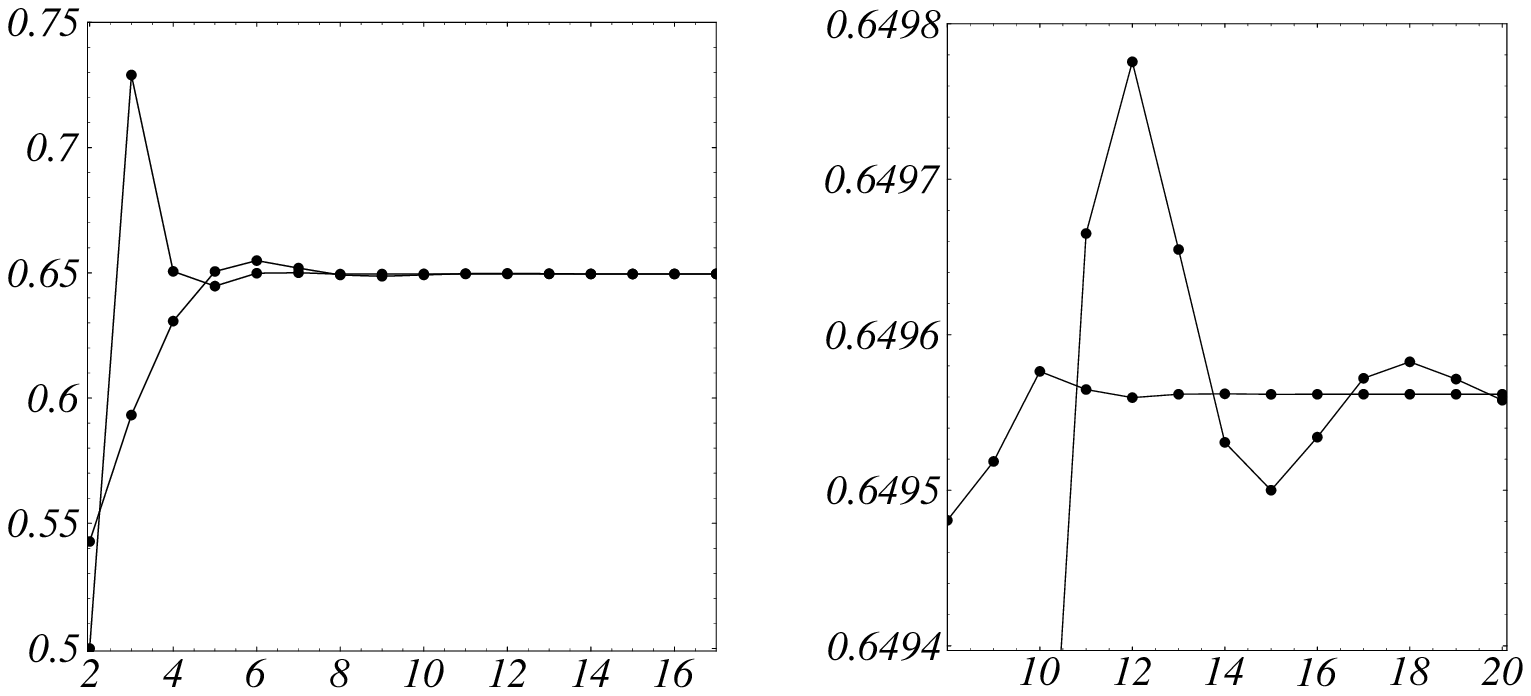,width=.8\hsize}
\end{picture}
\vskip-5cm
\begin{minipage}{.92\hsize}
  {\small {\bf Figure 10:} The critical index $\nu$ for the Ising
    universality class as a function of the order of the truncation
    $n_{\rm trunc}$, for expansion I around vanishing field and
    expansion II around the non-trivial minimum.  Left panel: I and II
    converge towards the asymptotic value.  Right panel: Magnification
    by a factor of 625.  The variation of all data points is below
    $10^{-3}$. II converges faster than I.  Expansion I (II)
    fluctuates about the asymptotic value with decreasing amplitude
    and a six-fold (four-fold) periodicity; see also Fig.~9.}
\end{minipage} 
\end{center}
\end{figure}

Until now, we have only employed expansion II, defined in
\eq{PolyAnsatz}. It is worthwhile to employ as well expansion I,
defined in \eq{PolyAnsatz0}. This has been done for the critical
exponent $\nu$ of the Ising universality class in Fig.~10. The left
panel shows that {\it both} expansions lead to a fast convergence
towards the asymptotic value.  The right panel is a magnification by
625, showing that expansion II indeed converges faster, although for
$n_{\rm trunc}=10$ their difference is already below $10^{-3}$.
Expansion I fluctuates about the asymptotic value with decreasing
amplitude and a six-fold periodicity in the pattern $+++---$.

\subsection{Convergence and scheme dependence} \label{SectionScheme}

In this section, we discuss the convergence of the ERG flows and the
polynomial expansion for various regulators. In Fig.~11, we have
computed the critical exponent $\nu$ $(N=1)$ for the sharp cutoff
$r_{\rm sharp}(y)=1/\,\theta(1-y)-1$, the quartic regulator $r_{\rm
  quart}(y)=y^{-2}$ and the optimised regulator \eq{ropt}. The left
(right) panel uses the expansion I (II).  Both $r_{\rm quart}$ and
$r_{\rm opt}$ are optimised regulators in the sense coined in section
\ref{Optimisation}. From Fig.~11, three results are
noteworthy. First, for the expansion I, we confirm that the
convergence is very poor for the sharp cutoff. For both the quartic
and the optimised regulator we find a good convergence.  Second, the
convergence is additionally improved by switching to the expansion II.
Third, the critical exponents obey $\nu_{{\rm sharp}}> \nu_{{\rm
    quart}}> \nu_{{\rm opt}}$. Hence, the better the convergence and
the stability of the flow, the smaller the resulting critical exponent
$\nu$. This observation is also linked to the convergence of the
derivative expansion \cite{Litim:2001dt}.

\begin{figure}
\begin{center}
\unitlength0.001\hsize
\begin{picture}(850,750)
\put(200,370){\large $n_{\rm trunc}$}
\put(610,370){\large $n_{\rm trunc}$}
\put(300,650){I}
\put(700,650){II}
\put(280,450){\fbox{\large $\nu_{\rm Ising}$}}
\psfig{file=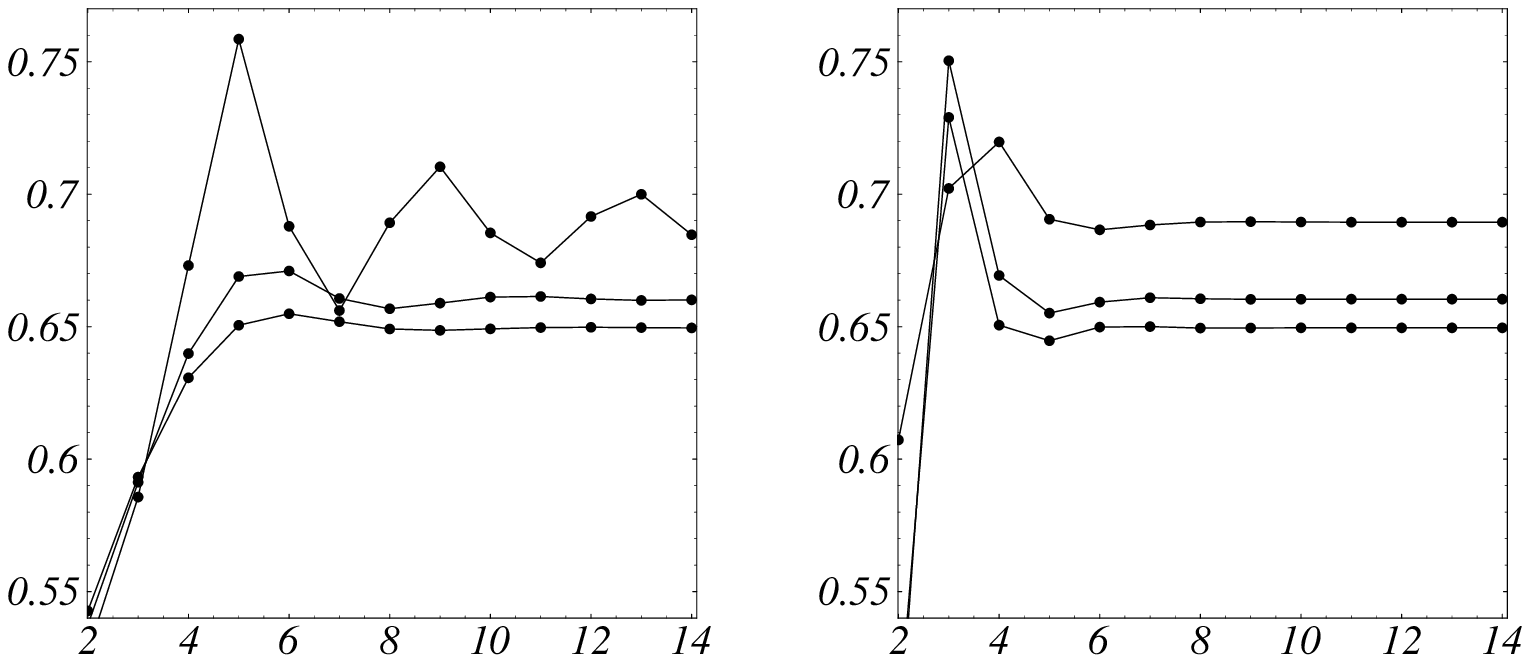,width=.8\hsize}
\end{picture}
\vskip-5cm
\begin{minipage}{.92\hsize}
  { \small {\bf Figure 11:} The critical index $\nu$ for the Ising
    universality class. Results are given for Expansion I (left panel)
    and II (right panel), and for the sharp cutoff (upper curves) the
    quartic regulator (middle curves) and the optimised regulator
    (lower curves).}
\end{minipage} \end{center}
\end{figure}

In \cite{Litim:2000ci}, it has been shown that the gap $C$ (the
radius of convergence for amplitude expansions) is linked to the
radius of convergence $C'$ for the expansions \eq{PolyAnsatz0} and
\eq{PolyAnsatz}.  An alternative way for identifying $C'$ consists in
studying the complex structure of the scaling solution. For real
$\rho$, the scaling solution is finite and real for all $\rho<\infty$.
In the complex plane, the scaling solution has (various) poles. Those
closest to the chosen expansion point constrain the radius of
convergence $C'$. For the sharp cutoff and the expansion I, this has
been analysed in \cite{Morris:1994ki} for $N=1$. It was found that
$\nu_{\rm sharp}$ cannot be determined to an accuracy better than
$8\cdot 10^{-3}$.  Hence, the polynomial expansion I for a sharp
cutoff does {not} converge beyond a certain level. \step

However, the findings of \cite{Morris:1994ki} do not imply that
polynomial truncations are not trustworthy {\it per se}. To the
contrary, the decisive difference between ``good'' or ``bad''
convergence properties stems essentially from an appropriate choice of
the regularisation. When optimised, the regularisation implies a
significant improvement for either expansion. In this light, the
non-convergence of the sharp cutoff flow within expansion I is
understood as a deficiency of the sharp cutoff regularisation. This
picture is consistent with the results of \cite{Aoki:1998um}, who
showed that expansion II leads to an improved convergence over
expansion I --- {even} for the sharp cutoff.

\section{Bounds on critical exponents} \label{Bounds}

In this section, we discuss how the critical exponents computed in the
previous sections, depend on the regularisation. This discussion is
mandatory because observables computed from a truncated flow, are
known to depend spuriously on the regularisation. It is decisive to
understand the range over which $\nu_{\rm ERG}$ may vary as a function
of the IR regulator. The origin of the spurious RS dependence is
easily understood. The regulator, while regulating the flow, also
modifies the coupling amongst all vertex functions of the theory.
These regulator induced contributions are of no relevance for the
integrated full flow, but they do matter for approximated flows, like
in the present case.  It is argued that the smallest value for the
exponent $\nu_{{}_{\rm ERG}}$ is obtained for the optimised regulator
$R_{\rm opt}$.  Prior to this, we recall the results obtained
previously in the literature, where, by a number of groups
\cite{Hasenfratz:1986dm,Morris:1994ki,Comellas:1997tf,Liao:2000sh,Litim:2001hk,Litim:2001dt},
critical exponents have been computed based on \eq{ERG} for different
regulators to leading order in the derivative expansion.  For all $N$,
all previously published results obey
\beq\label{range-lit}
\nu_{{\rm sharp}}  \ge  
\nu_{{}_{\rm ERG}} \ge
\nu_{{\rm min}}    \,. 
\eeq
The regulators studied in the literature cover the sharp cutoff and a
variety of smooth cutoff (exponential, power-law), and classes of
regulators interpolating between the sharp cutoff and specific smooth
cutoffs. Most results have been published for the Ising universality
class $N=1$.  For any $N\ge 0$, the smallest value $\nu_{{\rm min}}$
obtained in the literature is larger than the value $\nu_{\rm opt}$:
$\nu_{{\rm min}}>\nu_{\rm opt}$. For a detailed comparison of critical
exponents to leading and subleading order in the derivative expansion,
and a comparison to results from other methods and experiment, we
refer to Ref.~\cite{Litim:2001dt}.

\subsection{Upper boundary}

Now, we turn to a general discussion on the scheme dependence of
$\nu$. At first sight, \eq{range-lit} suggests that the possible range
for $\nu$ is bounded from above and from below. Let us assess the two
boundaries. We begin by showing that the inequality $\nu_{{\rm
    sharp}}\ge \nu_{{}_{\rm ERG}}$ does {not} hold for generic
regulator. Indeed, the upper boundary in \eq{range-lit} can be
overcome by choosing regulators which lead to a worse convergence than
the sharp cutoff. To see this more explicitly, consider a class of
regulators discussed in Appendix~\ref{VariantsOpt}. It is given by
\beq\label{Ra}
R_a(q^2)=a\,(k^2-q^2)\,\theta(k^2-q^2)\,,
\eeq
and is a variant of the optimised regulator \eq{Ropt}, to which it
reduces for $a=1$. For $a\to\infty$, it corresponds to the sharp cut
off. The regulator leads to an effective radius of convergence $C_a=a$
for $a<1$, and $C_a=\s0{a}{2a-1}$ for $a\ge 1$.  Since $C_a\ll 1$ for
$a\ll 1$, we expect that the corresponding critical exponents will
become large.  Our results for $R_a$ in \eq{Ra} are given in Fig.~12
(for more details, see Appendix~\ref{VariantsOpt} and Tab.~3). We find
that $\nu_a$ is a monotonously increasing function for decreasing
$a\le 1$.  In particular, for small $a$ we find indeed that
$\nu_a>\nu_{\rm sharp}$.  Hence, this result confirms the above
picture: Flows with a poor radius of convergence lead to large
numerical values for $\nu$.  \step

\begin{center}
\begin{tabular}{c||c|c|c|c|c|c|c}
$\gamma$&
$\s012$&
$\s034$&
$\s045$&
$\s056$&
1&
$\s032$&
2\\[.5ex] \hline
$\quad\nu_\gamma\quad$&
$\ \ \ 1\ \ \ {}$&
$\ .7354\ {}$&
$\ .7216\ {}$&
$\ .7142\ {}$&
$\ .6895\ {}$&
$\ .6604\ {}$&
$\ .6496\ {}$\\[-1ex]
\end{tabular}
\end{center}
\begin{center}
\begin{minipage}{.9\hsize}
  \vskip.3cm {\small {\bf Table 2:} Critical exponents $\nu$ (Ising
    universality class) for the flows $\ell_\gamma$ and various
    $\gamma$.\\}
\end{minipage}
\end{center}

\begin{figure}
\begin{center}
{}\vskip-2.cm
\unitlength0.001\hsize
\begin{picture}(1000,550)
\put(290,230){ \begin{tabular}{ccl}
$\ \ R_a:\ {}$      & $\ \bullet$\\[-.3ex]
$\ \ R_\gamma:\ {}$ & \ o \\[-.3ex] 
$\ \ R_c:\ {}$      & $\stackrel{\rm \ opt}{\put(0,0){\line(50,0){50}}}$
\end{tabular}}
\put(330,330){\fbox{{\Large $\displaystyle \nu_{{}_{\rm ERG}}$}}}
\put(260,-10){\large $x$}
\put(250,56){\small phys}
\put(230,85){\small opt}
\put(230,120){\small sharp}
\put(230,395){\small large-$N$}
\put(520,180){
\begin{minipage}{.47\hsize}
  { \small {\bf Figure 12:} Ising universality class.  The critical
    exponent $\nu$ for various classes of regulators.  For display
    purposes, we use $x=\s023(\gamma-\s012)$ for $R_\gamma$,
    $x=\0{a}{a+1}$ for $R_a$ and $x\equiv c$ for $R_c$. Boundaries:
    The full line (opt) corresponds to $R_c$ and denotes the lower
    boundary, the upper full line (large-$N$) denotes the upper
    boundary.  For comparison: The dashed line (sharp) indicates the
    sharp cut off value, and the thick full line (phys) the physical
    value.}
\end{minipage}} 
\hskip.02\hsize
\psfig{file=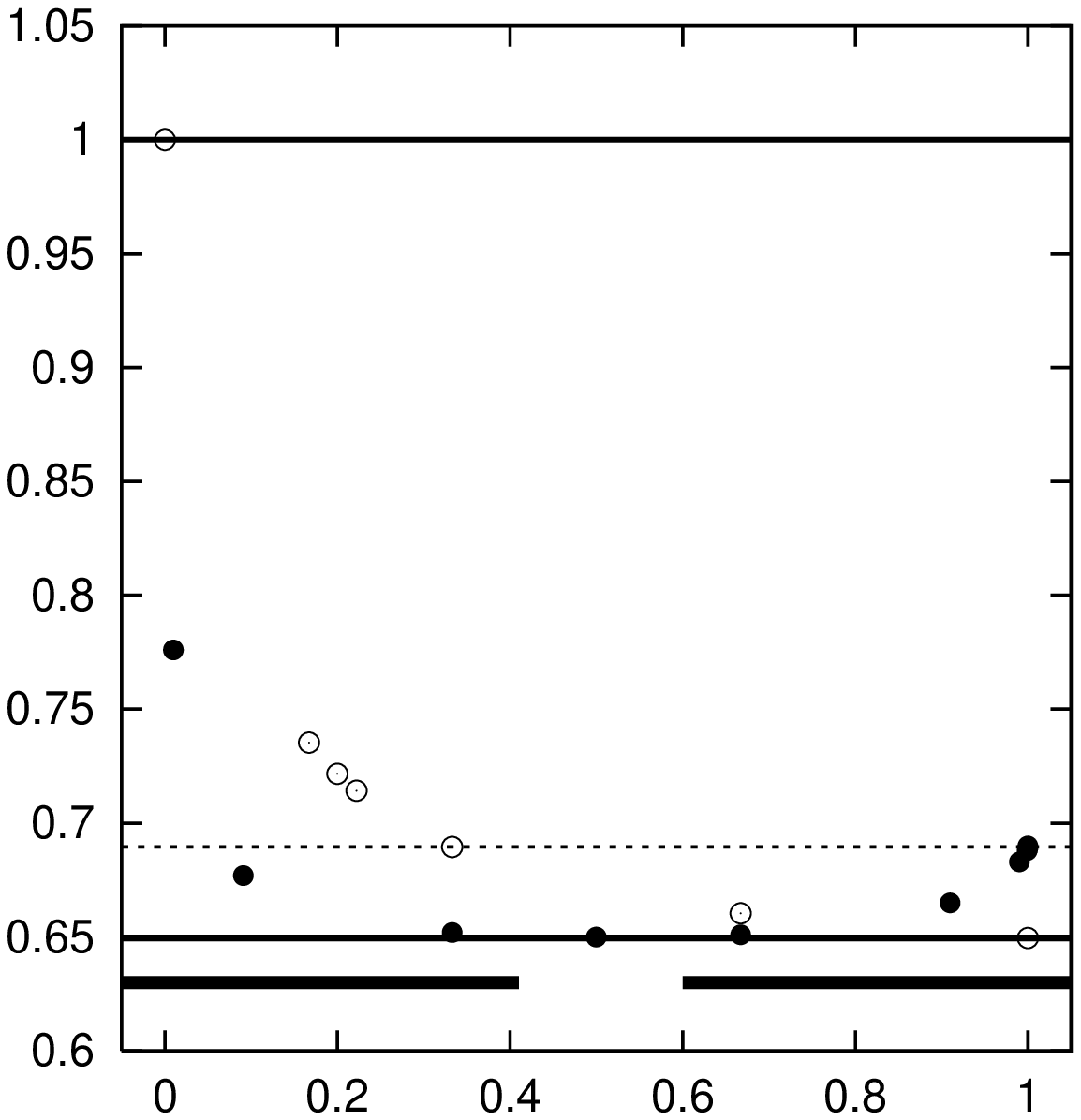,width=.45\hsize}
\end{picture}
\end{center}
\end{figure}

Next, we turn to a class of regulators aimed at probing the maximal
spread of flows. We recall that, to leading order in the derivative
expansion, all regulator-dependence is contained in the functions
$\ell(\omega)$. In Ref.~\cite{Litim:2001hk}, it has been shown that
the function $\ell(\omega)$ decays at most as $\omega^{-1}$, if the
regulator, obeying the basic constraints \eq{I}-\eq{III}, is not a
strongly oscillating function of momenta.  Hence, regulators leading
to a function
\beq\label{ell-gamma}
\ell_\gamma(\omega)\sim(1+\omega)^{1-\gamma}
\eeq
define for $\gamma=2$ a boundary in the space of regulators. This is
the case for $R_{\rm opt}$. No regulator can be found such that
$\gamma>2$ \cite{Litim:2001hk}. The proportionality constant in
\eq{ell-gamma} is irrelevant because it can be scaled away into the
fields in \eq{FlowPotential}. The second boundary is set by a mass
term regulator $R_{\rm mass}=k^2$: For a mass term, the corresponding
flow is a Callan-Symanzik flow which is not a Wilsonian flow in the
strict sense.  This comes about because the condition \eq{II} does no
longer hold true for $R_{\rm mass}$ in the limit $q^2\to\infty$. In
higher dimensions, this may lead to an insufficiency in the
integrating-out of large momentum modes.  Inserting $R_{\rm mass}$
into \eq{Id} in $d=3$ dimensions leads to \eq{ell-gamma} with
$\gamma=\s012$. This sets the second boundary.  The regulator
$R_\gamma$ is explicitly known for the cases $\gamma=2,\s032,1$ and
$\s012$, and corresponds, respectively, to the optimised regulator,
the quartic regulator $R_{\rm quart}=k^4/q^2$, the sharp cutoff and a
mass term regulator $R_{\rm mass}=k^2$.  For all other values of
$\gamma\in [\s012,2]$, $R_\gamma$ can be recovered explicitly from
$\ell_\gamma(\omega)$ \cite{Litim:2001hk}.  In the present case, we
only need to know that such regulators exist.\step

We have computed the critical index $\nu_\gamma$ for
$\ell_\gamma(\omega)$ with $\gamma\in[\s012,2]$, and our results are
given in Tab.~2. In Fig.~12, the results are denoted by open circles
for $x=\s023(\gamma-\s012)$.  As a result, the function $\nu_\gamma$
increases for decreasing $\gamma$, $\nu_\gamma\ge\nu_{\rm opt}$.  In
particular, once $\gamma<1$, the results obey $\nu_\gamma>\nu_{\rm
  sharp}$.  When $\gamma\to\s012$, the eigenvalues at criticality
approach their large-$N$ values $\nu\to 1$ and $\omega_n\to 2n-1$ for
any $N$. Note that the large-$N$ limit is exact in that it is
independent of the regularisation \cite{Litim:2001dt}. The large
numerical value for $\nu_{\gamma=1/2}$ is due to the deficiencies of
the Callan-Symanzik flow. We conjecture that the large-$N$ limit
$\nu_{{}_{\, {\rm large}\,N}}$ corresponds, for any $N$, to an upper
boundary for {any} regulator
\beq
\nu_{{}_{\rm  ERG}}\le \nu_{{}_{\, {\rm large}\,N}} = 1\,.
\eeq
%

\subsection{Lower boundary}

Next, we assess the lower boundary. It would be important to know whether
the optimised regulator leads to the smallest attainable value for
$\nu$ in the present approximation.  We are not aware of a general
proof for this statement. However, strong evidence is provided by
studying alterations of the optimised regulator. We have done so for
various classes of regulators, three of which are discussed here more
explicitly. For more details, we defer to the
Appendices~\ref{VariantsOpt} and \ref{SlidingOpt}. In
Appendix~\ref{VariantsOpt}, we employ variants of the optimised
regulator, given by the class $R_a$ of \eq{Ra}, and by the class $R_b$
defined as
\beq\label{Rb}
R_b(q^2)=
(k^2-q^2)\,\theta(k^2-q^2)\,\theta(q^2-\s012 k^2)
+(bk^2+(1-2b)q^2)\,\theta(\s012 k^2-q^2)\,.
\eeq
for $0\leq b\leq\infty$. This class contains the optimised regulator
\eq{Ropt} for $b=1$. The limit $b\to\infty$ corresponds to a variant
of the standard sharp cutoff.  Another new class of regulators $R_c$
is studied in Appendix~\ref{SlidingOpt}, where we allow for an
additional $k$-dependence on a mass scale within the regulator. We use
the class
\bea\label{Rc}
R_c(q^2) &=& 
(k^2-q^2-c\,m^2_k) \,\theta (k^2-q^2-c\,m^2_k)\,.
\eea
Here, $c$ is a free parameter and $m^2_k=U_k''(\phi=0)$ is the mass
term at vanishing field. For $c=0$, it turns into the optimised
regulator \eq{Ropt}. Due to the implicit scale dependence of $m_k$ on
$k$, the corresponding flow equations are substantially different from
the usual one. Most notably, they contain terms proportional to the
flow of $m_k$. \step

The classes $R_\gamma$, $R_a$, $R_b$ and $R_c$ probe ``orthogonal''
directions in the space of regulators.  $R_\gamma$ is sensitive to the
analyticity structure of the flow, $R_a$ and $R_b$ are sensitive to
alterations of the function $r(y)$ in the low momentum regime, and
$R_c$, while keeping the shape of the regulator $r(y)$ fixed, alters
the implicit modifications due to an additional running mass term.  As
such, $R_a$, $R_b$ and $R_c$ can be seen as variants of the optimised
regulator. The class $R_\gamma$ covers the largest domain of
qualitatively different flows \cite{Litim:2001ky}. 

\begin{figure}
\begin{center}
\unitlength0.001\hsize
\begin{picture}(1000,550)
\put(300,450){\fbox{{\Large $\displaystyle \0{\nu_{{}_{\rm ERG}}}{\nu_{\rm opt}}-1$}}}
\put(238,72){\large $N$}
\put(445,99){\large $\infty$}
\put(90,200){\small Wegner-Houghton}
\put(155,155){\small opt}
\put(210,300){\small Callan-Symanzik}
\put(520,200){
\begin{minipage}{.47\hsize}
  { \small {\bf Figure 13:} The spread of the critical exponent $\nu$ for
    various $N$ to leading order in the derivative expansion. The
    upper bound is set by the large-$N$ limit (Callan-Symanzik flow).
    The sharp cut-off results (Wegner-Houghton flow) are given for
    comparison.}
\end{minipage}} 
\hskip.02\hsize
\psfig{file=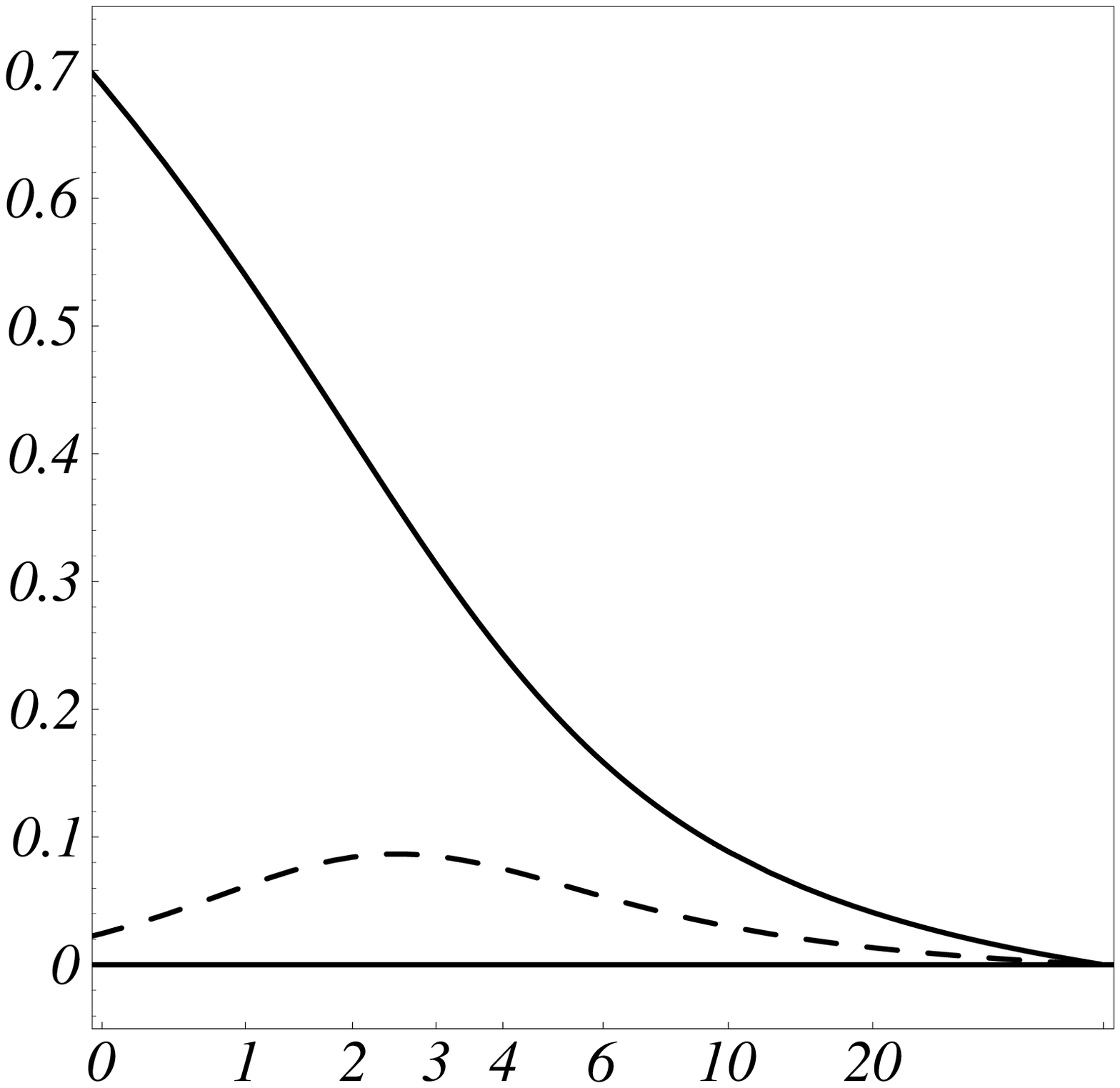,width=.45\hsize}
\end{picture}
\end{center}
\vskip-1cm
\end{figure}

We have computed the critical exponent $\nu$ for the Ising
universality class for all these classes of regulators. A part of our
results for $N=1$ is displayed in Fig.~12. Similar results are found
for all $N$. The classical (mean field) value for $\nu$ is $\nu_{\rm
  mf}=\s012$. The results for $R_a$ in \eq{Ra} are displayed in
Fig.~12 by full circles, with $x=\s0a{a+1}$. Some numerical values are
collected in Tab.~3 (cf.~Appendix~\ref{VariantsOpt}). It is found that
all critical indices $\nu_a$ obey $\nu_{{}_{\, {\rm large}\,N}}\ge
\nu_a\ge \nu_{\rm opt}>\nu_{\rm mf}$. This proves that alterations of
the regulator in the low momentum region do not lead to values for
$\nu$ smaller than $\nu_{\rm opt}$. An analogous result is found for
all regulators $R_b$ in \eq{Rb}, where $\nu_{{}_{\, {\rm
      large}\,N}}\ge\nu_{b}\ge \nu_{\rm opt}$. Some numerical values
are given in Tab.~4 (Appendix~\ref{VariantsOpt}).  In Fig.~12, our
results for $R_b$ are represented by open circles, and those for $R_c$
by a dashed line, with $x\equiv c$. All regulators $R_c$ from \Eq{Rc}
lead to the {\it same} critical exponent as $R_{\rm opt}$ in
\Eq{Ropt}, $\nu_c\equiv \nu_{\rm opt}$ (see
Appendix~\ref{SlidingOpt}).  \step

In Fig.~13, we discuss the spread of $\nu_{{}_{\rm ERG}}$ to leading
order in the derivative expansion for all $N\ge 0$. The spread
$\nu_{{}_{\,{\rm large}\,N}}/\nu_{\rm opt}-1$ is a $N$-dependent
quantity.  For $N=1$, the spread is about $0.54$, and hence quite
large. For comparison, the relative width with respect to the sharp
cut off $\nu_{\rm sharp}/\nu_{\rm opt}-1$ is roughly $0.06$ and
significantly smaller.  With increasing $N$, the spread vanishes as
$\sim 1/N$. This follows trivially from the fact that the ERG flow, to
leading order in the derivative expansion, becomes exact in the
large-$N$ limit \cite{Litim:2001dt}.
\step

In summary, the critical exponent $\nu$, as a function of the infrared
regularisation, is bounded. The upper boundary is realised for flows
with a mass term regulator, e.g.~Callan-Symanzik flows.  The lower
boundary is given by
\beq \label{range} 
\nu_{{}_{\rm ERG}} \ge \nu_{{\rm opt}}
\eeq
to leading order in the derivative expansion. Hence, $\nu_{{\rm opt}}$
appears indeed to be the smallest value attainable within the present
approximation. The inequality \eq{range} provides a quantitative basis
for the optimisation procedure which has lead to $R_{\rm opt}$ in the
first place. For the observable $\nu$, we have equally shown that the
optimised regulator corresponds, at least, to a ``local minimum'' in
the space of all regularisations. Furthermore, we have established
flat directions in the space of regulators. Based on the conceptual
reasons which have lead to $R_{\rm opt}$
\cite{Litim:2000ci,Litim:2001up}, we expect that $\nu_{\rm opt}$ even
corresponds to the global minimum.  At present, however, we have no
regulator-independent proof for this conjecture.

\section{Discussion and conclusions}\label{Discussion}

We studied in detail $O(N)$ symmetric scalar theories at criticality,
using the ERG method to leading order in the derivative expansion.
This included a complete investigation of the Gaussian fixed point in
$d>2$ for arbitrary regulator, and the computation of universal
critical exponents and subleading corrections at the Wilson-Fisher
fixed point for $d=3$.  Furthermore, we studied the spurious scheme
dependence for the critical exponent $\nu$ at the Wilson-Fisher fixed
point in three dimensions. One of the main new results is that the
leading critical exponent, as a function of the IR regulator, is
bounded from above and from below as
\beq\label{FinalBound}
\nu_{{\, {\rm opt}}} \le 
\nu_{{}_{\rm  ERG}}\le 
\nu_{{}_{\, {\rm large}\,N}}\,.
\eeq
This result has been achieved by studying the maximal domain of ERG
flows in the present approximation, ranging from Callan-Symanzik flows
to optimised flows and variants thereof.  The {qualitative} result --
the existence of a non-trivial Wilson-Fisher fixed point -- is very
stable, although the spread of values for $\nu$ is fairly large (see
Fig.~13). Supposedly, this is a consequence of a small anomalous
dimension $\eta$, which constrains higher order corrections. The
spread would shrink to zero only to sufficiently high order in the
derivative expansion or in the large-$N$ limit.\step  

The important {quantitative} question is: Which value for $\nu$ could
be considered as a good approximation to the physical theory? In view
of the regulator dependence, a prediction solely based on
\eq{FinalBound} is of little use. Our answer to this problem is
entirely based upon the structure of the ERG flow.  We proposed to use
specific regulators which lead to more stable ERG flows in the space
of all action functionals.  The numerical determination of critical
exponents and subleading corrections to scaling as given in Tab.~1, is
based on an optimised flow.  We expect that the results should be in
the vicinity of the physical theory.  In the present approximation,
the results for $\nu_{\rm opt}$ are indeed closest to the physical
ones \cite{Zinn-Justin:1989mi},
\beq\label{nu_Phys}
\nu_{{\rm  phys}}<\nu_{{\, {\rm opt}}}\,.
\eeq
The understanding of the spurious scheme dependence reduced the
ambiguity in $\nu$ to a small range about $\nu_{\rm opt}$. Typically,
the results from optimised flows other than $R_{\rm opt}$ are close to
the values achieved by $R_{\rm opt}$, and hence close to the lower
boundary of \eq{FinalBound}. In this light, optimised flows are
solutions to a ``minimum sensitivity condition'' in the space of all
IR regularisations \cite{Litim:2001fd}.  \step

Next we turn to the Callan-Symanzik flow, which is the flow with a
mass term regulator $R_{\rm mass}=k^2$. We argued that it defines the
upper boundary of values for $\nu$. It is quite remarkable that this
flow, for {any} $N$, leads to the same eigenvalues at criticality
given by the large-$N$ result.  Here, this result has been achieved
numerically. It would be helpful if it could also be understood
analytically. The large numerical value for $\nu$ reflects the poor
convergence properties of a Callan-Symanzik flow, essentially due to
deficiencies in the integrating-out of large momentum modes. \step

Now we discuss our results concerning polynomial approximations. The
reliability of this additional truncation is guaranteed if the
approximation convergences reasonably fast.  Here, we have established
that optimised flows converge very rapidly within the local potential
approximation. The efficiency is remarkable: a simple approximation
with only six independent operators --- say, the running v.e.v.~and
five running vertex functions up to $(\phi^a\phi_a)^5$ --- reproduces
the physical result for the exponent $\nu$ at the percent level.  A
better agreement cannot be expected, given that anomalous dimensions
of the order of a few percent have been neglected. These findings are
in contrast to earlier computations based on the sharp cut-off, where
the polynomial approximation has lead to spurious fixed point
solutions, even to high order in the approximation
\cite{Margaritis:1988hv}.  Hence, the efficiency of the formalism
not only depends on the choice for the degrees of freedom and the
truncation, but additionally, and strongly, on the IR regulator.  We
conclude that polynomial approximations are reliable for all technical
purposes, and even to low orders, if they are backed-up by appropriate
regulators. These considerations should be useful in more complex
theories whose algebraic complexity requires polynomial
approximations, e.g.~quantum gravity.\step

It would be interesting to apply the present ideas to theories like
QCD, where the propagating modes are strongly modified in the low
momentum regime due to confinement \cite{QCD1,Ellwanger:1996qf}. Then,
an optimised regulator is found by requiring that the regularised
inverse propagator is again flat, i.~e.~momentum-independent for small
momenta. Interesting choices are $R = (k^2-X)\,\theta(k^2-X)$ and $X =
\Gamma_k^{(2)}[\phi=\phi_0]$ and variants thereof. Here $\phi_0$
denotes a non-propagating background field. This conjecture is
supported by first results. \\[1ex]

{\bf Acknowledgments:}
I thank J.M.~Pawlowski for useful discussions and comments on the
manuscript. This work has been supported by the European Community
through the Marie-Curie fellowship HPMF-CT-1999-00404.\step

\setcounter{section}{0}
\renewcommand{\thesection}{\Alph{section}}
\renewcommand{\thesubsection}{\arabic{subsection}}
\renewcommand{\theequation}{\Alph{section}.\arabic{equation}}

\section{Gaussian fixed point}\label{Gauss}

In this appendix, we discuss the Gaussian fixed point of the flow
equation \eq{FlowPotentialOpt} in $d>2$ dimensions and for arbitrary
regulator. The Gaussian fixed point corresponds to the specific
solution $u_\star(\rho)={\rm const.}$ All higher derivatives of the
potential vanish, $u^{(n)}_\star(\rho)=0$. From the flow equation, we
deduce that
\beq\label{GFP:1}
u_\star=\0{2v_d}{d} N \int^\infty_0 dy\, y^{\s0d2 -1}\0{-r'(y)}{1+r(y)}\,.
\eeq
For the optimised regulator, we find $u_\star=4Nv_d/d^2$. More
information can be extracted by studying small perturbations $\delta
u^{(m)}$ around the $m$-th derivative of the scaling solution
$u_\star(\rho)^{(m)}$. The eigenperturbations obey the differential
equation
\beq\label{GFP:EW}
\partial_t\, \delta u^{(m)}=\omega\, \delta u^{(m)}
\eeq
with eigenvalues $\omega$. Expanding the flow equation to leading
order in $\delta u$, the eigenvalues obey
\beq\label{GFP:2}
0=
\left[\omega+d-(d-2)m\right]\, \delta u^{(m)} 
+\left[2A(\0N2+m)-(d-2)\rho\right]\,\delta u^{(m+1)}
+2\rho A \,\delta u^{(m+2)}\,.
\eeq
Here, the scheme-dependent coefficient $A$ is given by
\beq
A= 2v_d\int^\infty_0 dy \,y^{\s0d2 -3}\0{-r'(y)}{[1+r(y)]^2}
\eeq
and $0<A<\infty$. For the optimised regulator, $A_{\rm opt}=4v_d/d$.
Introducing new variables $x=(d-2)\rho/(2A)$ and $f(x)=\delta
u(\rho)$, the differential equation \eq{GFP:2} transforms into the
(generalised) Laguerre differential equation
\beq\label{GFP:Laguerre}
0=
\left(\0{d+\omega}{d-2}-m\right) f^{(m)}(x) 
+\left(\0N2 +m-x\right)f^{(m+1)}(x) 
+ x f^{(m+2)}(x)\,.
\eeq
We consider only polynomial solutions to \Eq{GFP:Laguerre}. The
requirement that solutions to \Eq{GFP:Laguerre} are bounded by
polynomials fixes the possible eigenvalues as
\beq\label{GFP:omega}
\omega=(d-2)(n+m)-d
\eeq
for non-negative integers $n$ and $m$. 
Apart
from an irrelevant normalisation constant, the $n$-th eigensolution to
\Eq{GFP:Laguerre} are given by the (generalised) Laguerre polynomials
\beq\label{GFP:LPoly}
\delta u^{(m)}(\rho)=
L^{m-1+N/2}_{n}\left(\0{2A\,\rho}{d-2}\right)\,.
\eeq
\Eq{GFP:LPoly} is the most general eigensolution at the Gaussian fixed
point in $d>2$ dimensions with eigenvalues given by \Eq{GFP:omega}.
The result holds for arbitrary regulator function. The scheme
dependence enters only the argument of the Laguerre polynomials in
\Eq{GFP:LPoly}. It is interesting to note that the eigenvalues are
independent of the regularisation scheme. Furthermore, the rescaled
differential equation \eq{GFP:Laguerre} is also independent of the
regulator.\step

In $d=3$ dimensions, the relevant and marginal operators are
$L^{N/2-1}_0$, $L^{N/2-1}_1$, $L^{N/2-1}_2$ and $L^{N/2-1}_3$ with
eigenvalues $\omega=-3,-2,-1$ and $0$ for $m=0$, or $L^{N/2}_0$,
$L^{N/2}_1$ and $L^{N/2}_2$ with eigenvalues $\omega=-2,-1$ and $0$
for $m=1$. In $d=4$ dimensions, the relevant and marginal operators
are $L^{N/2-1}_0$, $L^{N/2-1}_1$ and $L^{N/2-1}_2$ with eigenvalues
$\omega=-4,-2$ and $0$ for $m=0$, or $L^{N/2}_0$ and $L^{N/2}_1$ with
eigenvalues $\omega=-2$ and $0$ for $m=1$. For $m\ge \s0d{d-2}$, all
eigenoperators are marginal or irrelevant.\step

For $2m+N=1$ $(2m+N=3)$, the solution \Eq{GFP:LPoly} can be rewritten
in terms of Hermite polynomials of even (odd) degree,
\bea \label{GFP:m=0}
m=\s0{1-N}{2}:\quad \delta u^{(m)}(\rho)&=&
\0{(-1)^n}{n!}\,2^{-2n}\, 
H_{2n}\left(\sqrt{\0{2A\rho}{d-2}}\right)\\
\label{GFP:m=1}
m=\s0{3-N}{2}:\quad \delta u^{(m)}(\rho)&=&
\0{(-1)^n}{n!}\,2^{-2n-1}\, 
\sqrt{\0{d-2}{2A\rho}}\,
H_{2n+1}\left(\sqrt{\0{2A\rho}{d-2}}\right)\,.
\eea
Let us finally mention that some of these solutions have been given
earlier in the literature for the case of a sharp cutoff regulator
(with $A_{\rm sharp}=1$): for $N=1$ and $m=1$, \Eq{GFP:m=1} has been
given in Ref.~\cite{Hasenfratz:1986dm}, and \Eq{GFP:LPoly} for $m=1$
has been given in Ref.~\cite{Comellas:1997tf}.

\section{Variants of the optimised cutoff}\label{VariantsOpt}

In this appendix, we discuss variants of the optimised regulator
\eq{Ropt}. The aim is to probe whether certain alterations of the
regulator may lead to lower values for the critical exponent $\nu$.
Here, the properties of the regularisation are changed in the low
momentum region by modifying the function $r(y)$. In the following
appendix, we discuss modifications of $r(y)$ through the introduction
of additional (theory-dependent) $k$ dependent parameters.

\begin{figure}
\begin{center}
\vskip-1cm
\unitlength0.001\hsize
\begin{picture}(1000,550)
\put(250,250){\framebox{\large $y(1+r_a)$}}
\put(220,75){\large $y$}
\hskip.05\hsize
\put(520,200){
\begin{minipage}{.47\hsize}
  {\small {\bf Figure 14:} The function $y(1+r_a)$ for the class of
    regulators \eq{Radef}. All lines for different $a$ coincide for large
    momenta $y\in [1,\infty]$ (full line), but differ for small
    momenta $y\in [0,1[$ (dashed lines). The dashed lines clock-wise
    from the bottom: $a=\s012, 1, 2$ and $\infty$. The short dashed
    line corresponds to $a=0$ (no regulator). }
\end{minipage}} 
\psfig{file=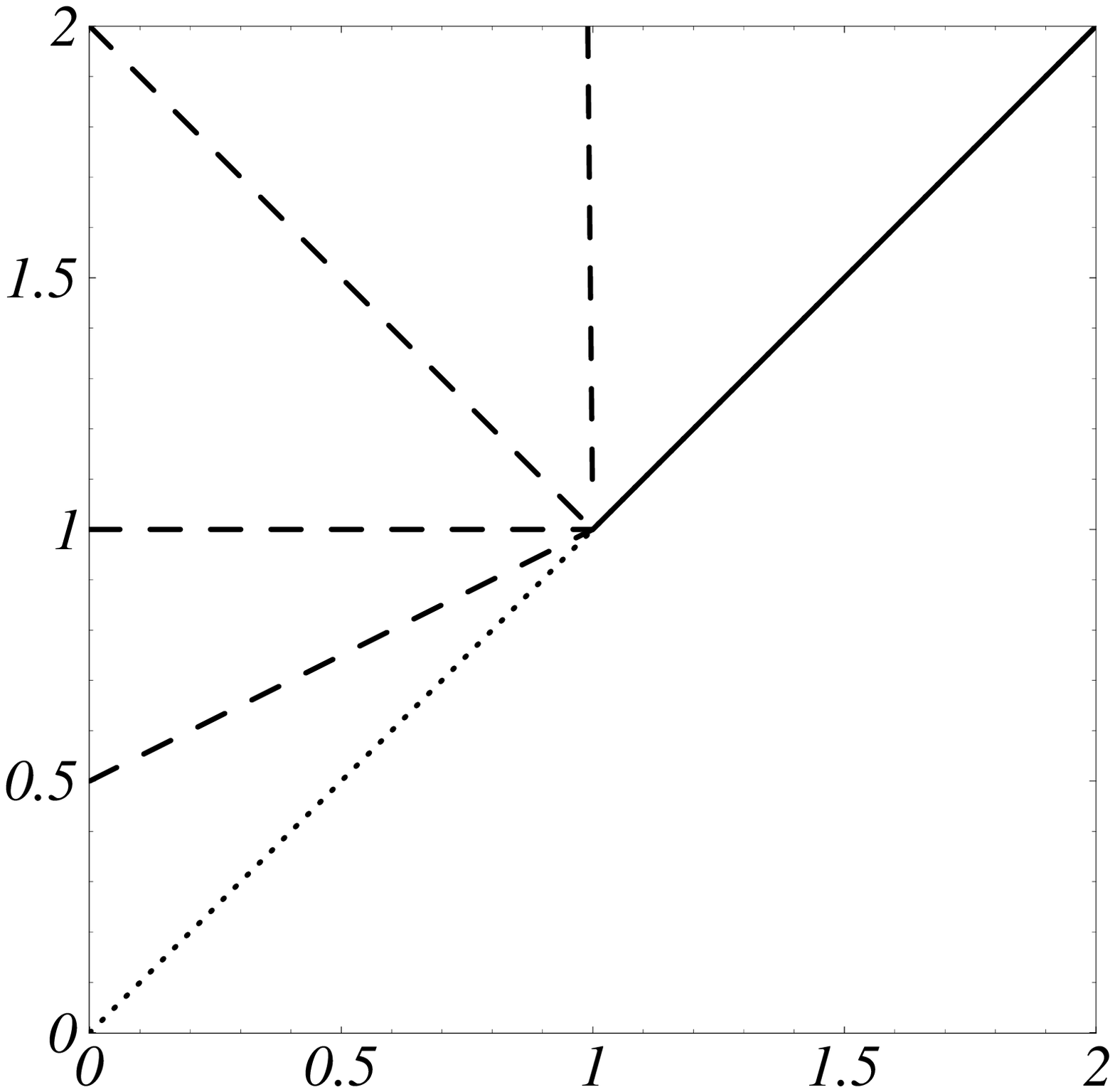,width=.45\hsize}
\end{picture}
\end{center}
\vskip-2cm
\end{figure}

\subsection{Definition}

In this appendix, we discuss two variants of the optimised regulator
\eq{Ropt}. First, we consider the class of regulators given by
\beq\label{Radef}
R_a(q^2)=a\,(k^2-q^2)\,\theta(k^2-q^2)\,.
\eeq
These regulators have a compact support. They vanish for all
$q^2>k^2$. In the infrared, they have a mass-like limit $R_a(q^2\to
0)= ak^2$ for all $0<a<\infty$. The limit $a\to\infty$ corresponds to
the sharp cut-off case. For $a=0$, the regulator is removed
completely. For $a=1$, the regulator \eq{Radef} reduces to the optimised
regulator \eq{Ropt}. As a function of $a$, the regulators differ only
in the momentum regime $y\in [0,1[\ $, where $y\equiv q^2/k^2$. The
dimensionless functions $r_a$ corresponding to \eq{Radef} are given by
\beq
r_a(y)=a\, (\s01y-1)\,\theta(1-y)\,.
\eeq
In Fig.~1, we have displayed the function $y(1+r_a)$ for various
cases. The full line corresponds to the range $y\in [1,\infty]$, for
all regulators \eq{Radef}. The dashed lines, clock-wise from the bottom,
correspond to $a=\s012, 1, 2$ and $\infty$.  The gaps associated to
\eq{Radef} are given by $C_a=\s0{a}{2a-1}$ for $a\ge 1$, and $C_a=a$ for
$\s012<a<1$. They are obtained from the normalised analogue of \eq{Radef},
chosen such that $r_a(\s012)=1$.
\step

Second, we consider another variant of the optimised regulator, where
the properties of the regularisation are changed only in the low
momentum region by modifying the function $r(y)$. Consider the class
of regulators given by
\beq\label{Rbdef} R_b(q^2)=
(k^2-q^2)\,\theta(k^2-q^2)\,\theta(q^2-\s012 k^2)
+(bk^2+(1-2b)q^2)\,\theta(\s012 k^2-q^2)\,.  \eeq
These regulators have a compact support. They vanish for all
$q^2>k^2$. In the infrared, they have a mass-like limit $R_b(q^2\to
0)= b\,k^2$ for all $0<b<\infty$. At first sight, it may seem that
\eq{Rbdef} is not a viable regulator for $b=0$, because $R_{b=0}$
vanishes in the infrared limit (no gap).  However, for $b=0$ the
function $\partial_t R_{b=0}$ vanishes identically for all $q^2<\s012
k^2$. Hence, \eq{Rbdef} provides a gap because
$R_{b=0}(\s012k^2)=\s012k^2>0$. For $b=1$, the regulator \eq{Rbdef}
reduces to the optimised regulator \eq{Ropt}. As a function of $b$,
the regulators differ only in the momentum regime $y\in [0,\s012[\ $,
where $y\equiv q^2/k^2$. The dimensionless functions $r_b$
corresponding to \eq{Rbdef} are given by
\beq
r_b(y)=
 (\s01y-1)\,\theta(1-y)\,\theta(y-\s012)
+(\s0by+1-2b)\,\theta(\s012 -y)\,.
\eeq
In Fig.~15, we have displayed the function $y(1+r_b)$ for various
cases. The full line corresponds to the range $y\in [\s012,\infty]$,
for all regulators \eq{Rbdef}. The dashed lines, clock-wise from the
bottom, correspond to $b=0, \s012, 1, 2$ and $\infty$. By
construction, the regulator is normalised as $r_b(\s012)=1$. The
associated gaps are given by $C_b=1$ for $b\ge 1$ and $b=0$ (see
below), and $C_b=b$ for $0<b<1$. For comparison, we have also given
the curve for the standard sharp cutoff (dotted line). The
corresponding gap is $C_{\rm sharp}=\s012$.  

\begin{figure}
\begin{center}
\vskip-.5cm
\unitlength0.001\hsize
\begin{picture}(1000,550)
\put(250,250){\framebox{\large $y(1+r_b)$}}
\put(220,75){\large $y$}
\hskip.05\hsize
\put(520,210){
\begin{minipage}{.47\hsize}
  {\small {\bf Figure 15:} The function $y(1+r_b)$ for the class of
    regulators \eq{Rbdef}. All lines for different $b$ coincide for
    large momenta $y\in [\s012,\infty]$ (full line), but differ for
    small momenta $y\in [0,\s012]$ (dashed lines). The dashed lines
    clock-wise from the bottom: $b=0, \s012, 1, 2$ and $\infty$. The
    standard sharp cutoff regulator (dotted line) is given for
    comparison. }
\end{minipage}} 
\psfig{file=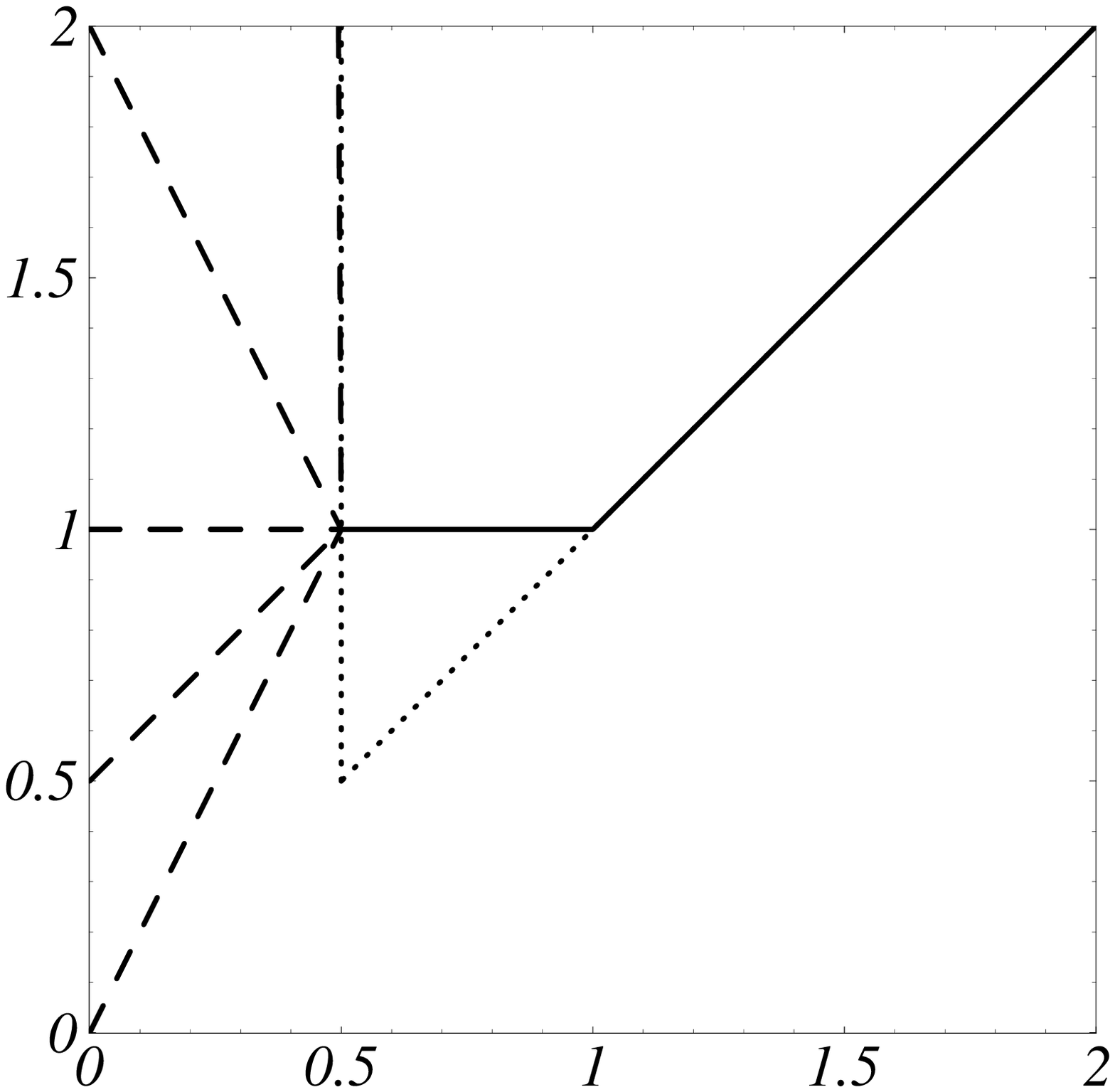,width=.45\hsize}
\end{picture}
\end{center}
{}\vskip-1.5cm
\end{figure}

The regulators $R_a$ and $R_b$ have a similar low momentum limit for
$q^2\to 0$, e.g.~$R_a(q^2\to 0)=R_{b=a}(q^2\to 0)=a\,k^2$. The crucial
difference between them concerns the intermediate momentum regime
$q^2\approx k^2$. Here, the regulator $R_b$ leads by construction to a
plateau for $y(1+r_b)$, which is absent for $R_a$.

\subsection{Flows}

In order to employ \eq{Radef} for the computation of critical exponents
in $d=3$, the associated flows $\ell(\omega)\to \ell_a(\omega)$ have
to be computed. For the function $\ell_a(\omega)$ and for $a\le 1$, we
find in $d=3$ (similar expressions are found for any $d$)
\beq
\label{la-}
\ell_a(\omega)=
\0{2a}{1-a}\left[1-\sqrt{\0{a+\omega}{{1-a}}}
\arctan\sqrt{\0{1-a}{a+\omega}}\right]\,.
\eeq
The region $a>1$ is obtained by analytical continuation:
\beq
\label{la+}
\ell_a(\omega)=
\0{2a}{1-a}\left[1-\sqrt{\0{{a+\omega}}{{a-1}}}
{\rm arctanh}\sqrt{\0{a-1}{a+\omega}}\right]\,.
\eeq
These flows have the following structure. They have poles on the
negative $\omega$-axis at $\omega=-1$. For $a=\infty$, the function
decays only logarithmically for asymptotically large $\omega$.  In the
limiting cases $a=1$ and $\infty$, we find
\bea
\label{la1}
\ell_{a=1}(\omega)&=& \s023 (1+\omega)^{-1}\\
\label{laInf}
\ell_{a=\infty}(\omega)&=&-\ln(1+\omega)+{\rm const.}
\eea
For $a=1$, in \eq{la1}, both expressions \eq{la-} and \eq{la+} have
the same limit discussed earlier in \cite{Litim:2001up}.  In the limit
$a\to\infty$, the resulting regulator is equivalent to the standard
sharp cut-off.
\step

In order to employ the regulator $R_b$ from \eq{Rbdef} for the
computation of critical exponents in $d=3$, the associated flows
$\ell(\omega)\to \ell_b(\omega)$ have to be computed. In full analogy
to the preceding computation, we find in $d=3$ (similar expressions
are found for any $d$) for the function $\ell_b(\omega)$ and for $b\le
1$
\beq
\label{lopta-}
\ell_b(\omega)=
\s023 (1-2^{-3/2})(1+\omega)^{-1}
+\0{b}{\sqrt{2}(1-b)}\left[1-\sqrt{\0{b+\omega}{{1-b}}}
\arctan\sqrt{\0{1-b}{b+\omega}}\right]\,.
\eeq
The region $b>1$ is obtained by analytical continuation:
\beq
\label{lopta+}
\ell_b(\omega)=
\s023 (1-2^{-3/2})(1+\omega)^{-1}
+\0{b}{\sqrt{2}(1-b)}\left[1-\sqrt{\0{{b+\omega}}{{b-1}}}
{\rm arctanh}\sqrt{\0{b-1}{b+\omega}}\right]\,.
\eeq
These flows have the following structure. They have poles on the
negative $\omega$-axis at $\omega=-1$ due to the first term on the
r.h.s.~in \eq{lopta-} and \eq{lopta+}. For large $\omega$ and $b<\infty$,
both expressions decay as $\omega^{-1}$. For $b=\infty$, the decay is
only logarithmic.  Let us consider the three limiting cases $a=0, 1$
and $\infty$. We find
\bea
\label{lopta0}
\ell_{b=0}(\omega)&=&\s023 (1-2^{-3/2})(1+\omega)^{-1}\\
\label{lopta1}
\ell_{b=1}(\omega)&=& \s023 (1+\omega)^{-1}\\
\label{loptaInf}
\ell_{b=\infty}(\omega)&=&\s023 (1-2^{-3/2})(1+\omega)^{-1}
-\s01{2\sqrt{2}}\ln(1+\omega)+{\rm const.}
\eea
For $b=0$, the momentum regime $q^2<\s012 k^2$ does not contribute to
the flow and the effective gap for $b=0$ is $C_0=1$. This is seen
directly from \eq{lopta-} and \eq{lopta0}: the first term of
\eq{lopta-} stems from the momentum interval $\s012 k^2\leq q^2\leq
k^2$, which is the only term surviving in \eq{lopta0}.  For $b=1$, in
\eq{lopta1}, both terms of \eq{lopta-} combine to the known result
discussed earlier in \cite{Litim:2001up}.  Finally, we turn to the
limit $b\to\infty$.  The resulting regulator is similar to the
standard sharp cut-off, with, however, an important difference.  For
the sharp cutoff, the function $y(1+r_b)$ has no plateau in the
momentum regime $\s012 k^2\leq q^2\leq k^2$ (see Fig.~15), which leads
to $\ell_{\rm sharp}(\omega)=-\ln(1+\omega)$. In \eq{laInf}, the
sharp-cutoff-like logarithmic term is clearly seen, and is due to the
momentum integration with $y\in [0,\s012]$.  However, a decisive
difference is the additional term in \eq{laInf}.  Notice also that the
constant in \eq{loptaInf} is actually infinite, but field independent.
Hence, it is irrelevant for a computation of critical exponents (only
the functions $\partial_\omega \ell_b(\omega)$ are needed).  \step

\subsection{Results}

For $R_a$, we have computed the critical exponent $\nu$ for the Ising
universality class using the flow equation \eq{FlowPotential} with
\eq{la-}, \eq{la+} (see Tab.~3). We confirm that
$\nu_{a=\infty}=\nu_{\rm sharp}$ and that $\nu_{a=1}=\nu_{\rm opt}$.
For all $a>1$ $(a<1)$, $\nu_a$ is a monotonically increasing
(decreasing) function with increasing $a$, hence $\nu_a\ge\nu_{\rm
  opt}$. Notice that the smallest value for $\nu$ is obtained for the
largest value for the gap parameter.  \step

\begin{center}
\begin{tabular}{c||c|c|c|c|c|c|c|c|c|c}
$a$&
$10^{-2}$&
$10^{-1}$&
$\s012$&
1&
2&
$10$&
$10^{2}$&
$10^{3}$&
$10^{4}$&
$\infty$\\[.5ex] \hline
$C_a$&
$10^{-2}$&
$10^{-1}$&
$\s012$&
1&
$\s023$&
$\s0{10}{19}$&
$\s0{100}{199}$&
$\s0{1000}{1999}$&
$\s0{10000}{19999}$&
$\s012$\\[.5ex] \hline
$\quad\nu_a\quad$&
$\ .776\ {}$&
$\ .677\ {}$&
$\ .652\ {}$&
$\ .650\ {}$&
$\ .651\ {}$&
$\ .665\ {}$&
$\ .683\ {}$&
$\ .688\ {}$&
$\ .689\ {}$&
$\ .690\ {}$\\
\end{tabular}
\end{center}
\begin{center}
\begin{minipage}{.95\hsize}
  \vskip.3cm {\small {\bf Table 3:} Critical exponents $\nu$ (Ising
    universality class) for the regulator $R_a$ and various $a$.}
\end{minipage}
\end{center}

\begin{center}
\begin{tabular}{c||c|c|c|c|c|c}
$b$&
$0$&
$\s012$&
$1$&
$10$&
$100$&
$\infty$
\\[.5ex] \hline
$\quad\nu_b\quad$&
$\ .6495\ {}$&
$\ .6518\ {}$&
$\ .6495\ {}$&
$\ .6594\ {}$&
$\ .6675\ {}$&
$\ .6699\ {}$\\
\end{tabular}
\end{center}
\begin{center}
\begin{minipage}{.95\hsize}
  \vskip.3cm {\small {\bf Table 4:} Critical exponents $\nu$ (Ising
    universality class) for the flows with $R_b$ and various
    $b$.}
\end{minipage}
\end{center}

For $R_b$, we have also computed the critical exponent $\nu$ for the
Ising universality class using the flow equation \eq{FlowPotential}
with \eq{lopta-}, \eq{lopta+}. We find that
$\nu_{b=0}=\nu_{b=1}=\nu_{\rm opt}$.  For $b<1$, we have
$\nu_b>\nu_{\rm opt}$. However, for too small values of $b$, $1\gg
b>0$, the regulator leads to a very small gap and the polynomial
approximation does no longer converge to a definite result, which is
an artifact of the regulator. For $b> 1$, $\nu_b$ is a monotonic
function of $b$ with $\nu_{b=\infty}\ge\nu_b>\nu_{\rm opt}$.  It is
interesting to discuss the case $b=\infty$ in more detail. Here,
$\nu_{b=\infty}=.6699\cdots$ which should be compared to $\nu_{\rm
  sharp}=.6895\cdots$ and to $\nu_{\rm opt}=.6495\cdots$. From the
structure of the flow, it is clear that the difference $\nu_{\rm
  sharp}-\nu_{b=\infty}$ has to be attributed to momenta with $y\in
[\s012,1]$. Hence, the ``flattening'' of the standard sharp cutoff
reduces the critical exponent by a few percent.  On the technical
level, this reduction is attributed to the non-logarithmic term in
\eq{loptaInf}. However, the smallest value for $\nu$ is obtained only
in case the logarithmic term is absent, as it happens in both
\eq{lopta0} and \eq{lopta1}.

\section{Cutoffs with intrinsic scaling}\label{SlidingOpt}

In this appendix, we study regulators with additional intrinsic
scale-dependent parameters.  

\subsection{Definition}

Up to now, we have considered IR regulators $R(q^2)$ which depend on
momenta only through the combination $q^2/k^2$, cf.~\Eq{r}.  In
\Eq{r}, the essential IR cutoff is provided by the function
$r(q^2/k^2)$, which cuts off the momentum scale $q^2$ in a way which
is independent of the particular theory studied. The $k$ dependence of
$R(q^2)$ is the trivial $k$ dependence linked to the dimensionality of
$R$. The situation is different once further $k$-dependent functions
are introduced into the regulator. Typically, this is done by
replacing
\beq\label{Rm} R(q^2) \to R(q^2,\{Z_k, m^2_k,\ldots\})\,. \eeq
Here, the set $\{Z_k, m^2_k,\ldots\}$ denotes scale-dependent
parameters of the specific theory studied, like the wave-function
renormalisation $Z_k$ or mass parameters $m_k$. It is expected that
the substitution \eq{Rm} leads to an improved convergence and
stability of the flow. A well-known example for \eq{Rm} is given by
$R(q^2) \to Z_k R(q^2)\,,$ which is often used beyond the leading
order in a derivative expansion. Here, the introduction of $Z_k$ in
the regulator simplifies the study of scaling solutions. In the
context of non-Abelian gauge theories, regulators like \Eq{Rm} have
been used in \cite{Ellwanger:1996qf} based on the replacement $q^2\to
\bar\Gamma^{(2)}_k[\phi_0]$ in the regulator, and hence
\beq\label{RGamma} R(q^2) \to  R[\bar\Gamma^{(2)}_k]\,.  \eeq
Again, this type of regulator has been motivated to stabilise the flow
and to encompass possible poles in the flow due to mass terms for the
gluonic fields \cite{Ellwanger:1996qf}. Notice that
$\bar\Gamma^{(2)}_k$ in \Eq{RGamma} cannot be the full field-dependent
functional, because elsewise the flow equation would no longer be the
correct one.  Instead, one has to evaluate it for some fixed
background field $\phi=\phi_0$ (hence the bar on
$\bar\Gamma^{(2)}_k$). In the present case, and to leading order in
the derivative expansion, we have
\beq\label{Gamma0}
\bar\Gamma_k^{(2)}[\phi_0]=q^2+U_k''(\phi_0)\,.
\eeq
Here, $U''(\phi_0)$ corresponds to a scale-dependent effective mass
term. Within the non-convex part of the potential, or in a phase with
spontaneous symmetry breaking, we have $m^2_k\equiv U''(\phi_0)<0$.
Hence the regulator \Eq{RGamma} is a special case of \Eq{Rm}. In the
remainder, the optimised regulator \eq{ropt} is used to define a class
of regulators of the form \eq{Rm}, namely
\bea\label{RMod}
R_c(q^2) &=& 
(k^2-q^2-c\,m^2_k) \,\theta (k^2-q^2-c\,m^2_k)\,.
\eea
Hence, $R_c(q^2)=R_{\rm opt}(q^2+c\,m_k^2)$. Effectively, this
corresponds to the replacement $k^2\to k^2_{\rm eff}(k)=k^2+c\,m^2_k$.
In terms of $r(y)$ defined in \Eq{r}, the regulator is given as
\bea
r_c(y) &=& (\s0{1-c\,\bar\omega}{y}-1) 
\,\theta (1-y-c\,\bar\omega)\,,
\eea
and the dimensionless mass parameter is
\beq \bar \omega \equiv m^2_k/k^2\,.\eeq
Below, we use $m^2_k\equiv U''(\phi_0=0)$. The regulator \eq{RMod} can
be seen as a `sliding' cutoff, because at any scale $k$, only the
$k$-dependent momentum interval $y\in [0,1-c\,\bar\omega_k]$
contributes to the flow. For $c=0$, the regulator reduces to
\eq{Ropt}, while for $c=1$ it turns into a regulator of the form
\eq{RGamma} with \eq{Gamma0}. Due to the additional dependence on
$m_k^2/k^2$, the structure of the flow equation is now different.

\subsection{Flows}

Let us compute the flow for the regulator \Eq{RMod}. Inserting
\Eq{RMod} into \Eq{Id}, and after some straightforward algebra, we
find
\beq\label{ell-mod} \ell_c(\omega)= \s02d (1-c\,\bar
\omega)^{d/2} \,
\0{1-c\,\bar\omega-\s0{c}{2}\partial_t\bar\omega}{1-c\,\bar
  \omega+\omega} \eeq
Notice that the function \Eq{ell-mod} depends now on $\bar \omega$ and
on the {\it flow} of $\bar \omega$. This is generic to
regulators of the form \eq{Rm} or \eq{RGamma}, because the implicit
scale dependence of $m^2_k$ leads to an additional term $\sim
\partial_t m^2_k$ in the flow equation. The flow equation
\eq{FlowPotential} with \Eq{ell-mod} becomes
\bea\label{FlowPotentialMod}
\partial_t u+du-(d-2) \rho u'
&=&
(N-1)(1-c\,\bar \omega)^{d/2} \, \0{1-c\,\bar
  \omega-\s0{c}{2} \partial_t \bar \omega}{1-c\,\bar
  \omega+u'}
\nonumber\\
&+&
(1-c\,\bar \omega)^{d/2} \, \0{1-c\,\bar
  \omega-\s0{c}{2} \partial_t \bar \omega}{1-c\,\bar
  \omega+u'+2\rho u''} \, \eea
Here, in order to simplify the expressions, we have rescaled the
irrelevant numerical factor $\s04d v_d$ into the fields and the
potential. In order to find an explicit form of the flow, the running
mass term needs to be specified. We chose $\bar \omega=u'(\rho=0)$.
This choice is motivated by the fact that the function $u'$, on the
fixed point, reaches its most negative value at vanishing field. For
$c=0$, the original flow may run into a pole at $u'=-1$. The
present choice shifts the pole to $u'(\rho)=-1+c\, u'(0)$. Since
$u'(\rho)-u'(0)\ge 0$ for a scaling solution, the right-hand side of
\eq{FlowPotentialMod} has no longer an explicit pole for $c=1$.
This has been the motivation for the structure of the regulator used
in \cite{Ellwanger:1996qf}. However, as we shall see, the full flow
still has an implicit pole due to the flow of $\bar \omega$ in
\eq{FlowPotentialMod}.  In terms of
\beq\label{MassScales}
\lambda_0\equiv u'(\rho=0)\,,\quad \lambda_1\equiv u''(\rho=0)\,,
\eeq
and after inserting \Eq{ell-mod} into \Eq{FlowPotentialMod} and
collecting terms proportional to the flow of $\lambda_0$, we find
\beq\label{flowmass} \partial_t \lambda_0
=-\0{2\lambda_0+\lambda_1{(N+2){(1-c\,\lambda_0)^{1+d/2}}}{
    [1+\lambda_0(1-c)]^{-2}}}{1-\s012 c\lambda_1
  {(N+2){(1-c\,\lambda_0)^{d/2}}}{
    [1+\lambda_0(1-c)]^{-2}}}\,.  \eeq
For the quartic coupling, and suppressing terms proportional to
$u'''(0)$, we find
\bea \partial_t\lambda_1 &=& -\lambda_1 \left(4-d- {\lambda_1}\,
  \frac{2(N+8)\,{\left( 1 - c\,\lambda_0 \right) }^{d/2}}{
    {\left[1 + (1-c)\lambda_0 \right]}^3} \right)\times \nonumber \\
\label{flowlambda}
&&\left( 1 - c\,\lambda_0 +\0{c}{2} \frac{\lambda_0
    +\0{\lambda_1}{2} {\left( N+2 \right){\left( 1 -c\,\lambda_0
        \right) }^{1+d/2}}{ {\left[ 1 + \left( 1 -c \right)
          \,\lambda_0 \right]^{-2} }}}{ 1 - c\0{\lambda_1}{2}
    {(N+2)\left( 1 - c\,\lambda_0 \right)^{d/2}}{\left[ 1 +
        \left( 1 - c \right) \,\lambda_0 \right]^{-2}}} \right)
\eea
The structure of the flows \eq{flowmass} and \eq{flowlambda} is easily
understood. The denominator in \eq{flowmass} stems from a resummation
of the back coupling of $\partial_t \lambda_0$. The denominator
becomes trivial for $c=0$. The numerator contains the usual scaling
term and a modified threshold behaviour, which depends now on $c$. A
similar structure appears for the flow of the quartic coupling.
Simpler forms of the flow are obtained for $c=0$ (no back-coupling of
a running mass term) or $c=1$.  For $c=0$, we have
\bea\label{flowmass0} \partial_t \lambda_0 &=&-2\lambda_0
+\lambda_1 \0{N+2}{(1+\lambda_0)^{2}}\,,\\
\label{flowlambda0}
\partial_t\lambda_1 &=& -(4-d)\lambda_1 + \lambda_1^2\,
\frac{2(N+8)}{(1 + \lambda_0 )^3}\,, 
\eea
while for $c=1$, the result is
\bea\label{flowmass1} \partial_t \lambda_0 &=& -\0{2\lambda_0
+\lambda_1{(N+2){(1-\lambda_0)^{1+d/2}}}}{1-\s012 \lambda_1
  {(N+2){(1-\lambda_0)^{d/2}}}} \,, \\
\label{flowlambda1}
\partial_t\lambda_1 &=& -\lambda_1 (1+ \lambda_0)\frac{4-d-
  {\lambda_1}\, {2(N+8)\,{\left( 1 -\lambda_0 \right)
      }^{d/2}}}{1-\s012\lambda_1 (N+2)(1-\lambda_0)^{d/2}}\,.  \eea
Hence, \eq{flowmass0} and \eq{flowlambda0} have a putative pole at
$\lambda_0=-1$. In turn, \eq{flowmass1} and \eq{flowlambda1} have
a putative pole at $(N+2)\lambda_1={2}(1-\lambda_0)^{-d/2}$. The
putative pole is absent for $N=-2$.

\subsection{Results}

From the explicit form of the function \eq{ell-mod}, and without
having made yet the explicit choice \eq{MassScales} for the mass term
$m^2_k$, we conclude that the non-universal fixed point solution
$\partial_t u=0$ of \eq{FlowPotential} for either $R_{\rm opt}$ from
\eq{Ropt}, or the class of regulators $R_c$ from \eq{RMod} are related
by a simple rescaling of the fields. The reason is the following: on
the fixed point $\partial_t u=0$, we also have $\partial_t
(m_k^2/k^2)=0$. Hence, the functions \Eq{ell-mod} do no longer depend
on the flow of the mass term. Hence, \Eq{ell-mod} becomes
$\ell(\omega)=\s02d C^{d/2+1}/(C+\omega)$, with $C\equiv 1-c\, \bar
\omega={\rm const}$ on a fixed point. This function is equivalent to a
flow derived from $R_{\rm opt}$. Hence, it suffices to rescale $u\to
u/C^{d/2}$ and $\rho\to\rho/C^{d/2+1}$ in order to transform the fixed
point solution for arbitrary $c$ onto the fixed point equation for
$c=0$.\step

Next, we check the $c$ dependence of critical exponents. A priori, the
critical exponents are sensitive to the flow in the vicinity of the
fixed point, and hence to the additional terms $\partial_t \lambda_0$
in the flow equation. We have solved the flow \eq{FlowPotentialMod}
with \eq{flowmass} numerically, within the polynomial approximation
defined in \Eq{PolyAnsatz0} up to $n_{\rm trunc} =20$. The eigenvalues
at criticality are found to be {\it independent} on the parameter $c$.
The results correspond to the lower full line in Fig.~12 (and $x\equiv
c$). Furthermore, we found that the critical exponents for different
$c$ agree for every single order in the truncation.  Hence, for $N=1$,
the exponent $\nu_{\rm trunc}$ as obtained from \eq{FlowPotentialMod}
with \eq{flowmass} are given by the line I in Fig.~10. \step

In summary, the introduction of scale-dependent parameters into the
regulator has lead to a significant change of the flow equations and
their pole structure. Hence, the approach to a fixed point solution,
and stability properties of the flow are quite different. Here, we
studied the replacement $R_{\rm opt}(q^2)\to R_{\rm
  opt}(q^2+c\,m^2_k)$. Universal quantities are independent of $c$,
because the modified regulator is linked to the optimised one through
the replacement $k^2\to k^2_{\rm eff}(k)=k^2+c\,m^2_k$ in the flow
equation.  This should be irrelevant for universal observables at a
fixed point, as has been confirmed explicitly. For the same reason,
the entire class of regulators $R_a(q^2+c\,m^2_k)\equiv a R_{\rm
  opt}(q^2+c\,m^2_k)$ leads to $c$-independent critical exponents.
Still, the flows are completely different for different $a$.
Therefore, $c$ can be used as a free parameter to stabilise the flow,
without affecting the physical result. If the substitution $R(q^2)\to
R(q^2+c\,m^2_k)$ cannot be rephrased as a redefinition of the infrared
scale, it is expected that also universal observables no longer remain
insensitive to free parameters like $c$. This case should be studied
separately.


\end{document}